\documentclass{aa}  
\usepackage{graphicx}
\usepackage{placeins}
\usepackage{txfonts}

\begin{document} 

   \title{Resolving the large exoKuiper belt of the HD~126062 debris disc and extended gas emission in its vicinity}
   \titlerunning{The exoKuiper belt of HD~126062 and extended gas in its vicinity}   
   \authorrunning{J. M. Miley et al.}

    \author{James M. Miley \inst{1,2,3,4,5}, Grant M. Kennedy\inst{6}, Álvaro Ribas\inst{7}, Enrique Macias\inst{8}, John Carpenter\inst{4}, Miguel Vioque\inst{8}, Kevin Luhman\inst{9,10}, Thomas Haworth\inst{11},  Philipp Weber\inst{1,2,3}, Sebastian Perez\inst{1,2,3}, Alice Zurlo\inst{2,12} }

   \institute{Departamento de Física, Universidad de Santiago de Chile, Avenida Victor Jara 3659, Santiago, Chile 
   \and Millennium Nucleus on Young Exoplanets and their Moons (YEMS), Chile 
   \and Center for Interdisciplinary Research in Astrophysics and Space Exploration (CIRAS), Universidad de Santiago, Chile
   \and Joint ALMA Observatory, Alonso de Córdova 3107, Vitacura, Santiago, Chile 
   \and European Southern Observatory, Alonso de Córdova 3107, Vitacura, Santiago, Chile
   \and Malaghan Institute of Medical Research, Gate 7, Victoria University, Kelburn Parade, Wellington, New Zealand
   \and Institute of Astronomy, University of Cambridge, Madingley Road, Cambridge CB3 0HA, UK
   \and European Southern Observatory, Karl-Schwarzschild-Str. 2, Garching bei München, Germany
   \and Department of Astronomy and Astrophysics, The Pennsylvania State University, University Park, PA 16802, USA
   \and Center for Exoplanets and Habitable Worlds, The Pennsylvania State University, University Park, PA 16802, USA
   \and Astronomy Unit, School of Physics and Astronomy, Queen Mary University of London, London E1 4NS, UK
   \and Instituto de Estudios Astrof\'isicos, Facultad de Ingenier\'ia y Ciencias, Universidad Diego Portales, Av. Ej\'ercito Libertador 441, Santiago, Chile}

   \date{}

  \abstract
   {Intermediate mass stars (1-3 M$_\odot$) host some of the brightest and most well-studied debris discs. This stellar class is also the most frequent host of molecular gas in systems with ages beyond typical protoplanetary disc lifetimes, and the most likely to host detected giant planets in radial velocity surveys. The debris discs of intermediate mass stars have therefore become a fertile ground for studying disc-planet interactions.}
   {In this work we present the first ALMA observations towards the A-type star HD~126062, located in Upper Centaurus Lupus/Lower Centaurus Crux, in order to characterise the properties of its debris disc.}
   {We probe the thermal continuum emission using observations at 1.3~mm, which are analysed through image reconstruction employing different visibility weighting regimes in addition to parametric model fitting to the observed visibilities. The observational setup also covers the frequency of the $^{12}$CO molecular line, allowing for imaging of gas in the vicinity of the system. }
   {We detect the dust continuum emission from an exoKuiper belt around HD~126062. Modelled as a Gaussian ring, the visibilities are consistent with a radial separation, R= 2.01$_{-0.05}^{+0.04}$$''$, equivalent to $\approx$270$^{+5}_{-4}$~au, and a full width half-maximum of $\Delta$R=0.71$''\pm0.09$, or 95$\pm 12$~au. The continuum emission appears in an almost face-on configuration with inclination to the line of sight constrained to be  $\leq17^\circ$. $^{12}$CO(2--1) emission is detected in the vicinity of the debris disc, with the majority of the emission found external to the exoKuiper belt. }
   {The exoKuiper belt characterised here is one of the largest to be detected, and is consistent with previous predictions of the distribution of dust in the system made from SED fitting. The morphology and displacement in velocity with respect to the systemic velocity suggest that the gas is not associated with the star and  debris disc, it most likely originates from a diffuse gas cloud in the nearby Galaxy.} 


   \maketitle
   
%

\section{Introduction}

Debris discs around A-type stars have been useful testing grounds for exploring the formation and evolution of planetary systems. Some of the first debris discs to have the structure of their thermal emission spatially resolved were A-type stars such as Beta Pictoris, Vega and Fomalhaut. This was initially achieved at sub-millimetre wavelengths by observations using SCUBA at JCMT \citep{Holland1998SubmillimetreStars}. Subsequent high resolution investigations into each of these systems would resolve and characterise their exo-Kuiper belts, which is the major component of the cold dust distribution \citep[]{MacGregor2017,Matra2019KuiperALMA,Matra2020DustALMA,Gaspar2023SpatiallyJWST/MIRI, Rebollido2024JWST-TSTMIRI}. Debris discs were traditionally assumed to be gas-poor environments, but now there are over 20 discs \citep{Marino2020PopulationStars}, including Beta Pic and Fomalhaut \citep{Dent2014MolecularDisk,Matra2017DetectionComets}, where significant amounts of CO gas have been detected. A large proportion of these gas-rich discs are hosted by A-type stars \citep{Moor2017}. As a result, debris discs around this particular spectral class have become a significant influence on our understanding of the origin of gas in debris discs \citep[e.g. see][]{Cataldi2023PrimordialALMA} and its kinematic behaviour as probed by the ALMA Large program ARKS (Marino et al. in prep). 

In addition to being key touchstones for defining the distribution of solid circumstellar material in evolved planetary systems, intermediate mass stars also play a key role in the detection of exoplanets; in radial velocity surveys the giant planet detection rate peaks at a host mass of $\approx 2.0$M$_\odot$ \citep{Johnson2010GiantPlane, Reffert2014}. 
Intermediate mass stars are the hosts to some of the best studied cases of confirmed planetary systems that reside within detectable debris discs. Systems such as these provide an excellent opportunity to directly study planet-disc interactions, particularly given the very low rate of detections of exoplanets embedded within the earlier protoplanetary disc phase. For example, Beta Pic has one confirmed exoplanet, Beta Pic b \citep{Lagrange2009AImaging} and evidence of another candidate planet, Beta Pic c, from HARPS RV data \citep{Lagrange2019EvidenceSystem}. A-type star HR~8799 hosts a very wide debris disc, reaching $\approx~$400~au, in a nearly face-on projection \citep{Booth2016ResolvingALMA,Faramaz2021A7}. Within its inner cavity are 4 giant planets \citep{Marois2008Direct8799., Marois2010Images8799}, and as such it has since become an important reference point for studies of giant planets and the debris discs within which they are embedded. 

We do not yet fully understand every step on the pathway from a gas-rich, planet-forming protoplanetary disc, to a planet-hosting system in which the natal disc has dispersed. This transition plays a large role in setting the future exo-planetary architecture \citep{Raymond2011DebrisFormation} and potentially in determining the atmospheric composition of planets within the system \citep{Kral2020FormationAccretion}. 
\citet{Pericaud2017} identify young gas-bearing debris discs that could be indicating a `hybrid disc' phase (again, most are A or F type stars), where dust appears similar to that in a debris disc (optically thin, as created through a collisional cascade), yet the gas is more similar to that found at the protoplanetary disc stage, in particular it is rich in H$_2$. The system HD~141569 has a protoplanetary-like gas mass yet its dust mass is more similar to debris discs, potentially indicating an intermediary stage of evolution \citep{Dent2005,Wyatt2015,White2016,Miley2018}. 
\citet{Wyatt2015} explore the main stages in the transition from protoplanetary disc to debris disc, and discuss the key physical processes at work. 
One of these key processes is the dispersal of molecular gas. Debris discs were traditionally considered to be gas-poor, but now gas detections have been made in relatively young debris discs, such as in the samples of \citet{Lieman-Sifry2016} and \citet{Moor2017} that belong to Upper Centaurus Lupus and Lower Centaurus Crux \citep[$\approx$ 15-20 Myr][]{Luhman2022AComplex}, as well as in much older systems \citep[$\eta$ Corvi, 1-2 Gyr,][]{Marino2017ALMAPlanets}. Both of these timescales are in tension with those typically expected for the photodissociation of CO in the ISM of 120 yr \citep{Visser2009} and the typical lifetimes of protoplanetary discs \citep[$\approx$ 10 Myr,][]{Ribas2014DiskMyr, Wyatt2015}. The quantities of gas detected in these systems would have significant effects on the nature of exoplanets present within the system should it be accreted, as \citep{Kral2020FormationAccretion} investigate in the case of terrestrial planets. As a result there is great interest in determining the true nature of this gas, and understanding how it has survived so long. The release of a secondary generation of gas though collisions between planetesimals in the debris belts may be able to explain the presence of gas at late stages \citep{Marino2020PopulationStars}, relying on atomic carbon to shield CO from rapid removal via photo-dissociation \citep[e.g.][]{Kral2019ImagingDiscs}, but current models overpredict the amount of atomic carbon (CI) compared to observations \citep{Cataldi2023PrimordialALMA}. Models of primordial gas, i.e. that which has somehow survived from the protoplanetary stage, appear to be more consistent with observed CI quantities \citep{Cataldi2023PrimordialALMA}, but still need to explain how the molecular gas has survived for so long. 

A-type stars may hold the key to investigating the long-lived gas in debris discs and in determining its origin. Modelling by \citet{Nakatani2021PhotoevaporationRemnants, Nakatani2023AStars} finds that depletion of small dust grains can reduce photoelectric heating, which in turn reduces FUV photoevaporation rates in evolved discs around A-type stars, potentially explaining how they retain gas discs at advanced ages, and lending support for a `primordial' gas model in which the gas remains from the protoplanetary disc stage. With a secondary gas model, in which gas is released from collisions between planetesimals, \citet{Marino2020PopulationStars} find that A-type stars are the most likely to birth massive planetesimal belts with the capacity to hold a sufficient quantity of frozen-out volatile gas that will later be released through collisions. Both studies outline the importance of initial conditions, in particular initial gas mass, in setting the mass of material that remains at late stages.

An ideal place to collect data on this transitional phase is in regions with ages just beyond the typical lifetimes of protoplanetary discs, in which the majority of discs are expected to have been dispersed or are in the act of being dispersed. A prime example is the region of Lower Centaurus Crux/Upper Centaurus Lupus (LCC/UCL), at an age of 15-20 Myrs \citep{Pecaut2016TheAssociation, Luhman2022AComplex}. Models of planetesimal collisions suggest the discs at this age should be high in luminosity of thermal dust emission, and if their collisions are enhanced by the stirring action of an inner exoplanet they may even be approaching their luminosity peak \citep{Wyatt2008EvolutionDisks}. This provides the perfect environment for studying evolved systems that are planet-hosting rather than planet-forming, through which we can study disc-planet interactions. Furthermore they may provide fertile grounds for finding molecular gas in debris discs; 7/11 of the gas bearing debris discs in the sample of \citet{Moor2017} were found in these regions. 

In this paper we present new ALMA observations of the circumstellar environment of A-type star HD~126062, found in the region of UCC/UCL \citep{Luhman2022AComplex}. The first evidence of a disc around the star was found with Spitzer photometry at 24 and 70~$\mu$m \citep{Chen2012AScorpius-Centaurus,Chen2014TheTargets}. Here we show the first characterisation of its debris disc at mm wavelengths.
We analyse the new data in order to characterise the distribution of dust in the HD~126062 debris disc and assess its potential for future studies of debris disc-exoplanet interactions. 
In the following section we describe the new observations, and in Section 3 we present the ALMA images and parametrise the dust density distribution through fits to the observed visibilities. In Section 4 we discuss these results and place them in the context of other known systems, in Section 5 we summarise our main conclusions.

\section{Observations }

HD~126062 is part of LCC/UCL \citep{Luhman2022AComplex}, its ICRS coordinates are 14h24m37.000s -47d10m39.864s and the star is located a distance of 134.1$^{+0.5}_{-0.6}$~pc \citep{Bailer-Jones2021Estimating3} as derived from Gaia DR3 parallaxes\citep{Brown2021iGaia/i3,Vallenari2023iGaia/i3}.
ALMA observations were taken using Band 6 receivers on 28th and 29th December 2022, on both occasions employing 46 antennas. During the first observation the average precipitable water vapour (pwv) at zenith was 0.54~mm and the median phase rms was measured at 8.6 $^{\circ}$ on the bandpass calibrator, using the 80th percentile projected baseline on a timescale equal to that of the phase referencing timescale. During the second observation pwv was 0.35~mm with a median phase rms was 9.5$^{\circ}$. The phase calibrator was J1407-4302 and the flux and bandpass calibrator was J1427-4206. The total integration time spent on the science target HD~126062 was 16 minutes. 
Two spectral windows, centred at 216.997 and 232.997~GHz, were dedicated to continuum observations, with a total bandwidth of 1.875 GHz and channel widths of 15625 kHz. One spectral window was centred on the $^{12}$CO (2--1) molecular transition, with a total of 3840 spectral channels of width 244 kHz, or 0.31~km/s. The final spectral window contained 1920 channels of width 977 kHz, or 1.27~km/s. This window covers both the $^{13}$CO (2--1) and C$^{18}$O (2--1) molecular lines, neither of which were detected. All line-free spectral channels are used to create the continuum images.

Continuum images were reconstructed using CASA's tclean task with a Hogbom minor cycle deconvolution algorithm, \citep{Hogbom1974ApertureBaselines}. Images were created using a range of weighting schemes, in this article the fiducial image is constructed using natural weighting in order to prioritise signal-to-noise ratio over angular resolution. To scrutinise extended structures in the image, we also create images after applying a uvtaper to the visibilities of size 1$\farcs$5 . The naturally weighted continuum image achieves a synthesizing beam with FWHM $0\farcs95\times0\farcs85$ and position angle $ 87^{\circ}$, with an rms noise of 25~$\mu$Jy/beam. When the uvtaper is applied to the data the resulting synthesized beam has size $1\farcs85\times1\farcs65, 90^{\circ}$ and the rms is measured at 40~$\mu$Jy/beam.
Images of the $^{12}$CO (2--1) transition are created by re-gridding the data along the spectral axis into channels of width 0.62~km/s using a rest frequency of 230.5238~GHz, before applying tclean with natural weighting. We apply masks to the channels using the \textsc{auto-multithresh} algorithm implemented in CASA's tclean task \citep{Kepley2020Auto-multithresh:Algorithm}. This achieves a synthesizing beam of $0\farcs95\times0\farcs83, -84.7^{\circ}$, producing an image cube with an rms noise level measured at 2~mJy/beam within each channel. The moment 0 map of the $^{12}$CO emission is created by summing all pixels that contain emission above the 3$\sigma$ level in individual channels.

\begin{figure*}[h]
    \centering
    \includegraphics[width=0.99\linewidth]{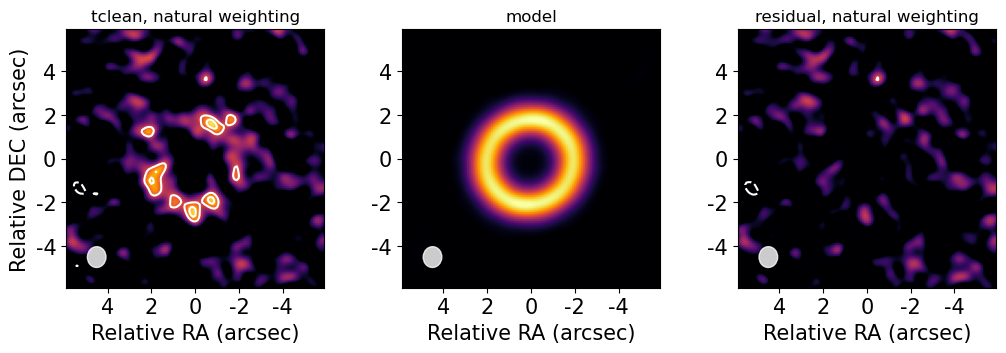}  
  
    \includegraphics[width=0.99\linewidth]{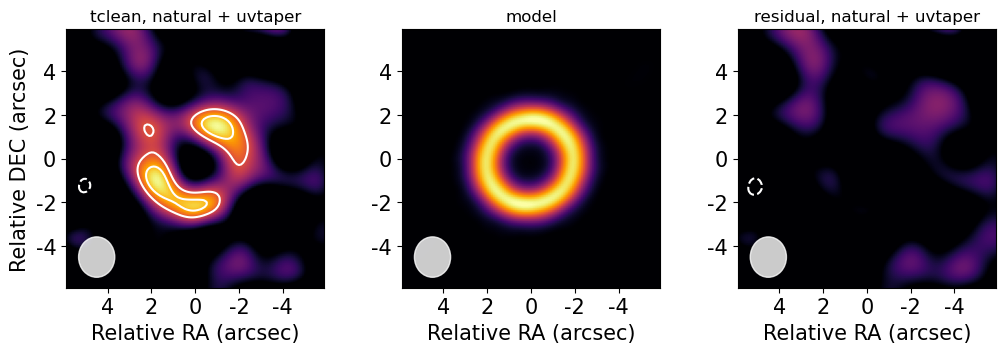}
    
    \caption{\textbf{Top row:} On the left is the fiducial image of the continuum observations of HD~126062 reconstructed using tclean with natural weighting. In the middle is the Gaussian ring model derived from the MCMC fit to the visibilities. On the right is the residual image following the subtraction of the model from the data, restored using the same imaging parameters as the tclean image.    \textbf{Bottom row:} Same as the top row, except tclean images are reconstructed with a uvtaper of 1$\farcs$5 applied. Contours are drawn in the tclean and residual images at -3,3,4 $\times \sigma_{\rm im}$, where $\sigma_{\rm nat}= $ 25~$\mu$Jy/beam and $\sigma_{\rm taper}= $ 40~$\mu$Jy/beam. Negative contours are shown as dashed lines.}
    \label{fig:tclean_images}
\end{figure*}

\section{Results}

\subsection{An exoKuiper belt detected at 1.3~mm}
\label{sec:res_alma}

Through continuum imaging we detect an extended belt of mm emission around the star for the first time.
Many images of the dataset were created by varying the visibility weighting in order to scrutinise the faint emission. Figure \ref{fig:tclean_images} shows two representative examples from the imaging experiments; the top left panel was created with natural weighting, whilst the bottom left panel shows the image created with the additional uvtaper applied. In the natural weighted image the observed peak brightness is 114~$\mu$Jy/beam resulting in a peak signal-to-noise ratio of 4.6. The emission takes the form of several emission peaks at $\approx 4 \sigma$ that are distributed in a ring-like morphology that is more apparent in the tapered image. There are breaks in the ring-like brightness distribution at $\sim$58$^\circ$ and 238$^\circ$  measured East of North. The peak brightness of the tapered image is measured at 187 $\mu$Jy/beam.

To accurately quantify the basic geometric properties of the newly discovered ring of mm emission around HD~126062, we fit the visibilities with a parametric model using the \textsc{galario} package \citep{Tazzari2016} and uvplot \citep{Tazzari2017Mtazzari/uvplot}. Similarly to \citet[][]{Marino2016}, we assume that the mm dust distribution of the debris disc can be modelled as a Gaussian ring defined by 

\begin{equation}
    F_\nu = f_0 ~\rm{exp}~ \large[- \frac{1}{2} \Big(\frac{R_0 -R }{\sigma}\Big)^2 \large], \sigma= \frac{\Delta R}{2\sqrt{2\rm{ln}2}} , 
    \label{eqn:gaussRing}
\end{equation}

\noindent where $f_0$ is a flux normalisation factor, $R$ is the distance from the star, R$_0$ the centre of the ring and the width is parametrised by the standard deviation of the Gaussian, $\sigma$, which can be related to the full width at half maximum, $\Delta R$. We also sample the inclination (`inc') and position angle (`PA'). Using MCMC sampling through the emcee package \citep{Foreman-Mackey2013EmceeHammer} to sample the posterior probability distribution in order to achieve a representative model given the observed deprojected visibilities. For the results presented here, 50 walkers were used taking 18000 steps with a 1000 step burn-in phase. The effective sample size and autocorrelation time were calculated in order to ensure these chosen parameters were appropriate, and convergence was assessed by calculating the rank normalised R-hat diagnostic \citep{Gelman1992InferenceSequences}. Table \ref{tab:mcmcm} gives the adopted prior distributions for each of the properties of interest.

\begin{table}[]
\caption{Parameter space explored by the MCMC, uniform priors are adopted between the range of values indicated. }\label{tab:mcmcm}
\centering
\begin{tabular}{l|ll}
\hline
\multicolumn{1}{c|}{Parameter} & \multicolumn{2}{c}{Range}    \\ \hline
f0                             & \multicolumn{1}{l|}{1} & 10  \\
R (")                          & \multicolumn{1}{l|}{1} & 3   \\
$\Delta$R (")                  & \multicolumn{1}{l|}{0} & 6   \\
inclination (deg)                      & \multicolumn{1}{l|}{0} & 90  \\
PA (deg)                       & \multicolumn{1}{l|}{0} & 180 \\ \hline
\end{tabular}
\end{table}

We take a representative model from the MCMC by taking the median value from probability distribution. Diagnostic plots are shown in Appendix \ref{sec:vis_modelfit} in Figure \ref{fig:vis_fit}, which demonstrate that the model follows the binned visibilities well until they become dominated by noise at baselines larger than $\approx 200$k$\lambda$. The corner plot of distributions of each of the parameters is also shown in the bottom panel of Figure \ref{fig:vis_fit}. Precise results are found for the flux, the radial position and the width of the ring. Considering the distance to HD~126062 from Earth, the centre of the mm-ring is found to be at a physical separation from the star of $\approx270^{+5}_{-4}$~au. We can convert the standard deviation of the Gaussian ring into a physical width, which corresponds to a full width half maximum as shown by the expression for $\sigma$ in Equation \ref{eqn:gaussRing}, giving  $\Delta R$= 95$\pm$12~au.
The uncertainties on the physical radial separation and width of the ring are calculated by taking the 16th and 84th quartiles of the posterior distribution and converting these to a spatial scale considering the distance from Earth to HD~126062.
The MCMC results place an upper limit (where the limit is set by the 84th percentile) on the inclination of the exoKuiper belt of 17$^\circ$, with the lowest values preferred by the posterior distribution. We can conclude therefore that the data at hand are consistent with an almost face-on orientation. A face-on orientation naturally makes it difficult to constrain a position angle for the source, which can explain why this value remains unconstrained by the results in Figure \ref{fig:vis_fit}. Observations with higher angular resolution could place a more concise constraint on this value, which also depends upon the vertical height of the emitting material \citep[][]{Marino2016}, and would also reveal the sharpness of the inner edge of the debris disc, which itself can give indications about exoplanets present within the system \citep{Marino2020InsightsALMA}. Despite the areas of decreased brightness in the ring, which are clearest in the tapered image, the residual maps in Figure \ref{fig:tclean_images} demonstrate that the data is consistent with a symmetric brightness distribution. 
The HD~126062 exoKuiper belt has a large radius as should be expected if we assume a correlation between belt distance and stellar luminosity \citep[e.g. as in ][]{Matra2019KuiperALMA}, given that the central star is quite luminous (15.5 L$_\odot$). The belt of HD~126062 is, however, relatively thin. Its fractional width $\Delta R/R = 0.35\pm0.06$, whereas 70\% of REASONS survey discs are found to be wide $\Delta R/R > 0.5$ \citep{Matra2025REsolvedWavelengths}. The dust distribution around HD~126062 is discussed in more detail in Section \ref{sec:dust_dist}. 

To estimate the flux density from the exoKuiper belt, we measure a flux density from within a radius of $R + \Delta R$ as constrained by the results from the MCMC in uv-space as described above. In the natural-weighted image this results in a flux density measurement of 0.94$\pm$0.09 mJy/beam, where the uncertainty is calculated by propagating a 10\% uncertainty in ALMA flux calibration \citep{Cortes2025ALMAHandbook} and the statistical uncertainty in the measurement over this region. 

\subsection{Spectral energy distribution of HD~126062}
\label{sec:sed_fit}

\begin{figure}
    \centering
    \includegraphics[width=1\linewidth]{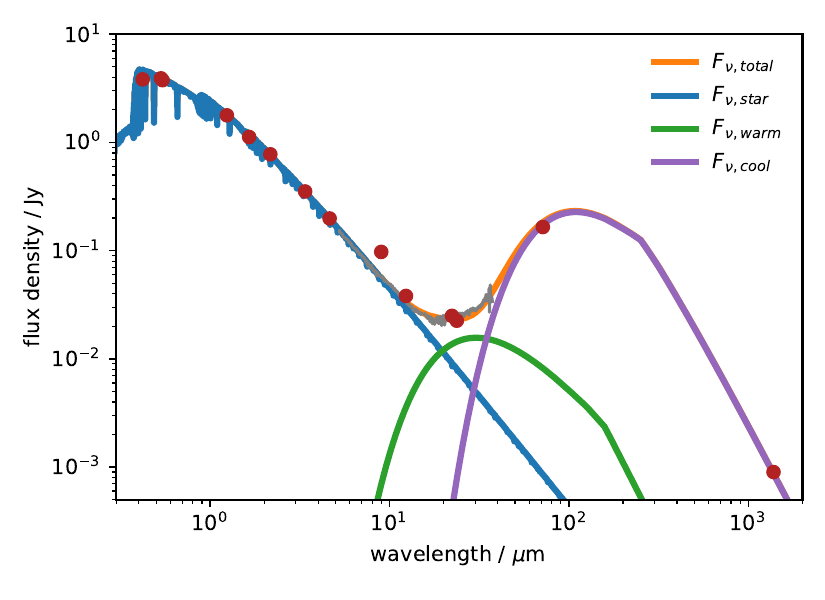}
    \caption{Fit to the spectral energy distribution of HD~126062, comprised of two blackbody components and a stellar contribution.\ Error bars are included, but are smaller than the data point markers.}
    \label{fig:sed}
\end{figure}

The only prior knowledge of a debris disc around HD 126062 comes from SED fitting. Debris discs are often well described by a combination of two blackbodies at different temperatures \citep{Kennedy2014DoBelts}, which can be interpreted as contributions from belts of material in the disc as distinct temperatures. Fitting the observed SED of debris discs can therefore give some indication of the distribution of dust density in the disc, \citep[as applied to HD~126062 in][]{Chen2014TheTargets,Launhardt2020ISPY-NACOStars,Pearce2022PlanetSurveys}. The properties of the exoKuiper belt predicted by previous SED fitting attempts are very similar to those we independently constrain here through the resolved mm detection. For example the inferred location of the outer belt in HD~126062 by \citet{Pearce2022PlanetSurveys} of 300$\pm$80~au matches very well to that which we constraint here through fits to observed visibilities in Section \ref{sec:res_alma} of R=$270^{+5}_{-4}$~au and $\Delta R= 95\pm12$~au.

Data taken at longer (i.e. $\approx$mm) wavelengths provide important constraints that aid these efforts, as they trace the larger grain sizes. Here we update the SED fit to HD~126062 by including the new photometric point at 1.3~mm of 0.94$\pm$0.16 mJy in order to place tighter constraints on location of grains emitting within the inner regions of the disc. 
We follow the procedure as outlined in \citet{Yelverton2019ASystems} giving results as shown in Figure \ref{fig:sed}. Two components are required to achieve a good fit; one characterised by a blackbody of temperature T$_{\rm warm}$= 168$\pm8$~K, and the other at T$_{\rm cool}$= 47$\pm$2~K. These temperatures can be interpreted in terms of distance from the central star by calculating the blackbody radius using
\begin{equation}
R_{\text{BB}} = 1 \, \text{au} \left( \frac{T_{\text{BB}}}{278 \,\text{K}} \right)^{-2} \left( \frac{L_*}{L_{\odot}} \right)^{1/2}
\end{equation}
The blackbody radius derived from the fit to the SED alone does not reflect an accurate disc location. To find this we require the application of some correction factor which takes into account dust grain emission properties \citep{Booth2013,Pawellek2015TheDiscs, Yang2024FirstStar}. Here we use a correction factor, $\Gamma$, to convert the derived blackbody radius into a more accurate disc location, which takes the form
\begin{equation}
R_{\rm SED} \equiv \Gamma R_{BB} {,} \quad  \Gamma = A \left( \frac{L_*}{L_{\odot}} \right)^B .
\end{equation} 
In calculating the conversion factor we adopt the approach of studies that analyse belts resolved at mm wavelengths by ALMA (as opposed to similar Herschel studies), taking A=2.92$\pm$0.50, B=-0.13$\pm$0.07 \citep{Pawellek2021AGroup,Pearce2022PlanetSurveys}. This results in a $\Gamma=2.0\pm0.5$.
The cool blackbody contribution from the SED fit therefore predicts a belt location at 286$\pm$79~au, where the error is propagated from uncertainties on the temperature of the blackbody and of the conversion factor $\Gamma$. This is consistent with the location constrained from the model fit to observed visibilities in the previous section of R=$270^{+5}_{-4}$~au. 

Debris discs with a single, wide belt can have similar SEDs to those with two spatially separated belts \citep{Kennedy2014DoBelts}. In this case the exo-Kuiper belt in HD~126062 appears to be relatively narrow (Section \ref{sec:res_alma}, $\Delta R/R = 0.35\pm0.06$), and so we can have more confidence in interpreting the SED fit as evidence of two dust components rather than one, wider belt. 
The warm blackbody contribution to the SED fit therefore predicts the location of an inner, warmer component to be at 22$\pm$6~au.
With this fit we also constrain properties of the central star, for which a best fit is found with T$_*$ = 8940 $\pm$ 60~K, L$_*$  = 15.5 $\pm$ 0.2 L$_\odot$, R$_*$  = 1.6 $\pm$ 0.1 R$_\odot$. We achieve tighter constraints on the position of the continuum emission in the outer disc in comparison to that presented by \citet{Pearce2022PlanetSurveys} where they find a belt location of 300$\pm$80 au. This is because here we use the newly constrained values for the stellar luminosity L$_*$, and the effective temperature, T$_{\rm eff}$ of the cool outer belt, derived from SED fitting including the new mm photometric point. 

\subsection{$^{12}$CO (2--1) emission in the vicinity of HD~126062}
\label{res:gas}

\begin{figure}
    \centering
    \includegraphics[width=0.99\linewidth]{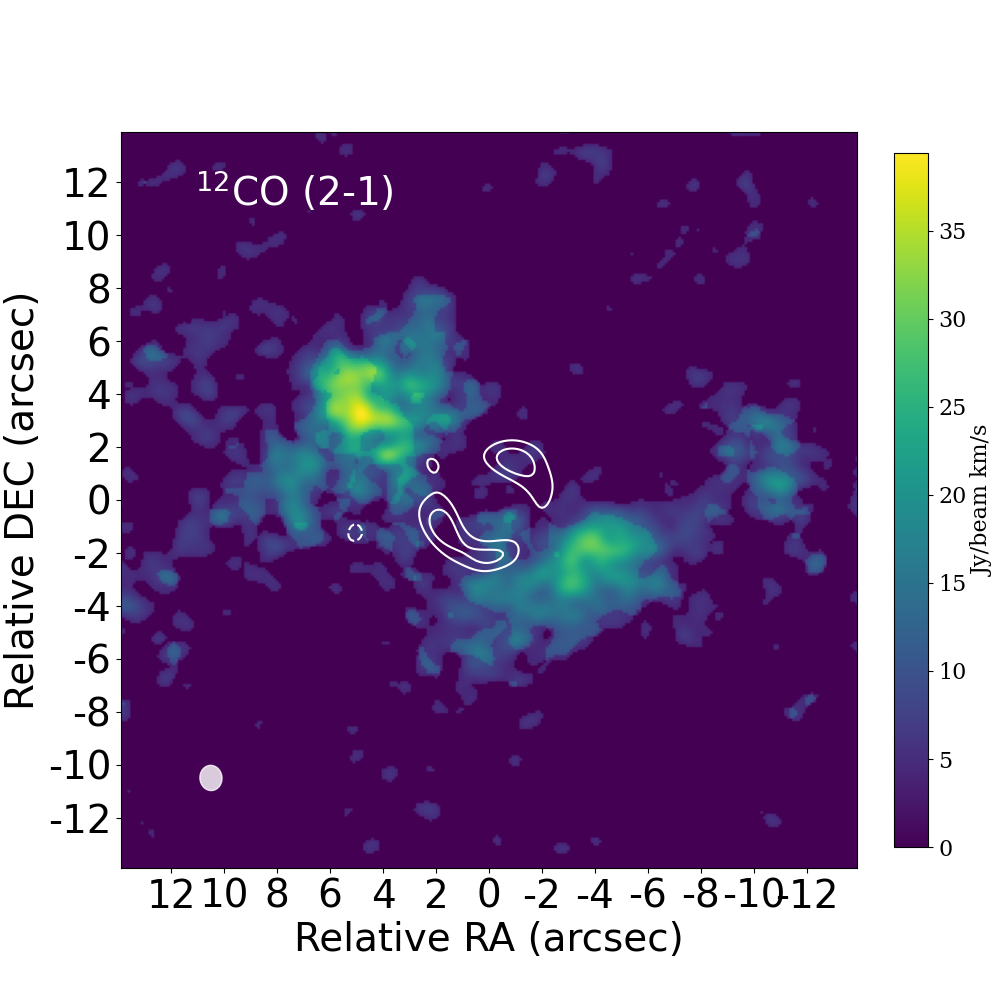}
    \caption{The integrated intensity (moment-0) map of the $^{12}$CO(2--1) transition observed towards HD~126062. Emission above $3\sigma$ in the channels is used to construct this moment map. White contours show the continuum image constructed with a uvtaper. The beam size is indicated in the lower left corner. }
    \label{fig:mom0}
\end{figure}

The observing setup of the ALMA observations covered the (2--1) transition of the $^{12}$CO molecule. In Figure \ref{fig:mom0} we present the moment 0 map of the emission detected toward HD~126062. The moment 0 map was created by integrating emission from channels with emission greater than $3\sigma$, each of which is also individually plotted in Appendix \ref{sec:appendix12CO}. Although $^{13}$CO(2--1) and C$^{18}$O(2--1) lines are covered by the spectral setup, they are not detected in these observations. 

\begin{figure}
    \centering
    \includegraphics[width=0.95\linewidth]{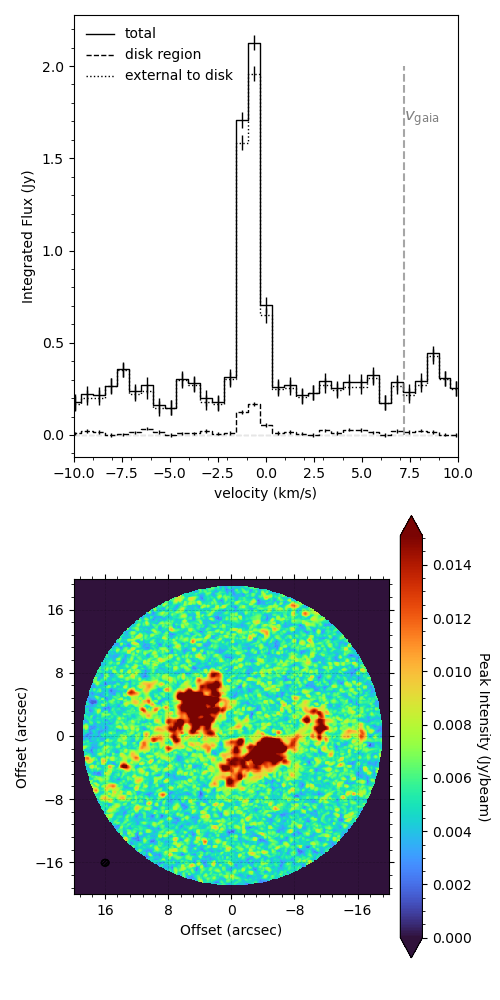}
    \caption{\textbf{Top:} $^{12}$CO \textbf{spectra} extracted from different regions of the image. Disk region is found within R+$\Delta R$, the external region begins at R+$\Delta R$ and extends as far as 3$\sigma$ emission is detected in the moment 0 map ($\approx 12^{"}$). Total encompasses both these regions. The systemic velocity as measured by gaia is indicated by a vertical line. \textbf{Bottom:} The moment 8 map of $^{12}$CO  emission, which collapses the cube in the spectral axis and shows the peak emission for each pixel in the cube.}
    \label{fig:spectra_and_mom8}
\end{figure}

The gas in Figure \ref{fig:mom0} shows an unusual morphology, with two lobes of emission appearing on either side of the dust disc along a NE-SW axis. Very little emission is detected from regions that are radially interior to the exoKuiper belt that is detected by the continuum observations. The rms is measured from a moment 0 image constructed without a noise clip, which gives a value of 4.1 mJy/beam km/s, the peak emission in the image reaches 9.6$\sigma$. 
In Figure \ref{fig:spectra_and_mom8} we show the spectrum of the $^{12}$CO line extracted from the detected emission. We also extract emission from specific regions in order to make a comparison. These additional regions are that which contains the exoKuiper belt detected in the continuum data, and also only the regions external to the disk, up to a radius of $12^"$, which encompasses the vast majority of the 3$\sigma$ emission in the moment 0 map (although some diffuse emission exists beyond this point, e.g. at relative RA=-12~$"$, relative DEC=0~$"$ in Figure \ref{fig:mom0}). Also plotted in the bottom panel of Figure \ref{fig:spectra_and_mom8} is the peak intensity along the line of sight for each pixel within the cube with a large field of view to show the surrounding environment. The $^{12}$CO is extended over large angular scales, 3$\sigma$ emission spans over 24$"$ in the field of view. The broad distribution of the emission appears roughly centred towards the position of the star, which invites speculation that the $^{12}$CO(2--1) emission we detect is associated with the system.   
The environment of HD~126062 in LCC/UCL has an estimated age of 15-20~Myrs, meaning one would generally expect for any cloud or envelope material is to have already been dispersed.

The gas line is displaced from the rest frequency of $^{12}$CO  by $\approx$1-2 km/s, and from the Gaia DR3 velocity for HD~126062 of 7.2~km/s (Figure \ref{fig:spectra_and_mom8}). 
The detected emission spans just three channels of size 0.62~km/s. A narrow line is expected for a rotating, nearly face-on disc, but the observed spatial distribution of the $^{12}$CO emission does not match this interpretation (See Figure \ref{fig:mom0}). A small amount of emission is detected within the disk region, i.e. that within the outer edge of the dust emission R$+\Delta R$, but its spectrum is very similar to that of the wider gas emission (as shown in Figure \ref{fig:spectra_and_mom8}), and therefore appears to be of a similar origin. The origin of the detected $^{12}$CO is discussed further in Section \ref{sec:discussionGas}.

\section{Discussion}

\subsection{Constraining the distribution of dust in HD~126062}
\label{sec:dust_dist}

\subsubsection{Millimetre radial brightness variation}

\begin{figure}
    \centering
    \includegraphics[width=0.98\linewidth]{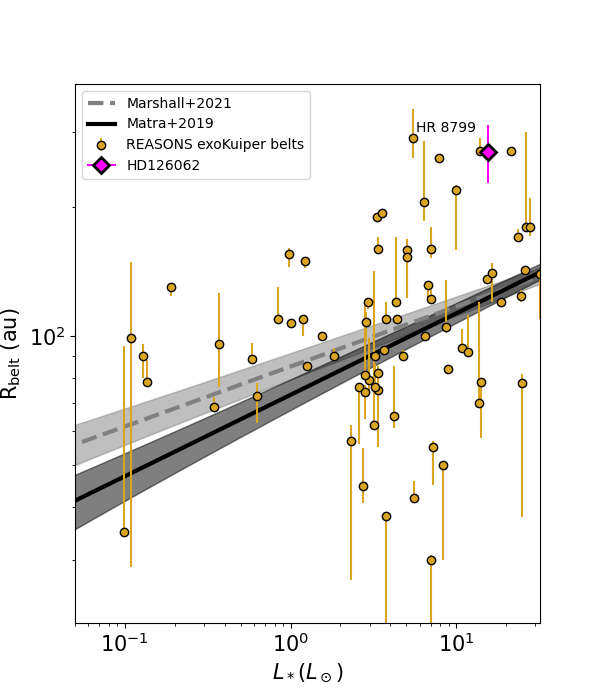}
    \caption{exoKuiper belt positions as characterised by the REASONS survey, plotted as a function of the stellar luminosity of their host, where error bars indicate $\Delta R$ as derived by the given references. HD~126062 is also plotted, using the stellar luminosity as derived from the SED fitting in this work. We also plot distance-stellar luminosity relations, $R \propto L^{\alpha}$ fit to samples of discs observed in the mm \citep{Matra2019KuiperALMA} and far-IR \citep{Marshall2021AWavelengths}.}
    \label{fig:exoKuipers}
\end{figure}

The exoKuiper belt of HD~126062 presented here with a radius of $\sim$270~au is one of the largest that has been detected to date. In Figure \ref{fig:exoKuipers} we plot the belt location in HD~126062 amongst the other known and spatially resolved exoKuiper belts from the REASONS survey \citep{Matra2025REsolvedWavelengths}. The most extended, angularly resolved exoKuiper belt is that of HR~8799 \citep{Booth2016ResolvingALMA,Faramaz2021A7}, a system known for having four confirmed giant planets orbiting interior to the belt at R=$290\pm10$~au. Other extended belts with a radius comparable to that of HD~126062 within its uncertainties also have a similar stellar luminosity. The most similar sources, which occupy the upper right of Figure \ref{fig:exoKuipers} are HD~84870 with R$_{\rm belt}$ = 260$\pm$50~au, HD~15257 has R$_{\rm belt}$ = $270^{+60}_{-40}$~au and HD~158352 has R$_{\rm belt}$ = 270$\pm$20~au, where the belt radii are taken from \citet{Matra2025REsolvedWavelengths}. All of these systems with very wide belts are A/F type stars and reside in the top right corner of Figure \ref{fig:exoKuipers}. Thus far only HR~8799 has been confirmed to host exoplanets. The size of this new exoKuiper belt contributes to maintain the previously identified trend of the brightest host stars possessing the exoKuiper belts with the largest radii. Both \citet{Matra2019KuiperALMA}, using data from ALMA at mm wavelengths, and \citet{Marshall2021AWavelengths} with far-IR data from Herschel, have fitted relations of the form $R_{\rm belt} \propto L_*^{\alpha}$. Although a large spread is identified in both studies, a trend of larger belts around more luminous stars does emerge. The REASONS sample of 74 planetesimal belts agrees with these previous results in identifying a general trend, although they find a shallower slope and a much larger spread in belt size for a given stellar luminosity. Selection bias towards more luminous targets may play a role here. Choosing only the most luminous discs will result in a shallower slope, and could explain the absence of very large belts around stars with lower luminosity \citep{Marshall2021AWavelengths,Matra2025REsolvedWavelengths}. Nevertheless, there still remains a lack of small belts of $\approx$ 10s of au in observations, which may indicate they are dispersed more quickly by collisional evolution \citep{Matra2025REsolvedWavelengths} in comparison to larger belts such as that around HD~126062.

\subsubsection{Axisymmetric brightness in the exoKuiper belt}

As demonstrated by the visibility fits to the continuuum observations described in Section \ref{sec:dust_dist}, the exoKuiper belt is consistent with an azimuthally symmetric ring. 
However in the images displayed in Figure \ref{fig:tclean_images}, the azimuthal brightness of the ring is not continuous. Two breaks in emission are particularly clear in the tapered image, with fainter parts of the rings found at  58$^\circ$ and 238$^\circ$. The detection of the ring material is in the 3-4$\sigma$ range, and so either noise or slightly fainter regions of the disc may explain the azimuthal variations.  
There are published cases of azimuthal brightness variations in debris discs; usually such features have been detected in scattered light observations or molecular gas observations, but are less pronounced or undetected in observations of thermal emission at mm wavelengths \citep[e.g. AU Mic, $\beta$ Pic][]{MacGregor2013MillimeterDisk,Matra2019KuiperALMA}, and so it may be that shorter wavelength observations reveal substructure asymmetry in HD~126062. Debris disc asymmetries can occur due to apocentric or pericentric `glow', where the exact nature of these features depends on the eccentricity of the disc \citep{Lynch2022EccentricDiscs}, as well as the width of the belt, the angular resolution of the observations \citep{Lovell2023EccentricDiscs} or the observing wavelength as demonstrated, for example, by the difference between Herschel and ALMA images of the Fomalhaut debris disc \citep{Acke2012HerschelActivity,MacGregor2017}.   
There are also comparisons that can be made with younger systems. The 180$^\circ$ separation between the two fainter parts of the ring is reminiscent of similar structures observed in observations of protoplanetary discs by ALMA \citep[e.g. J1604-2130, DoAr44,][]{Mayama2018ALMADisk,Arce-Tord2023Radio-continuum44} and in scattered light observations by SPHERE \citep{Bohn2022ProbingObservations}. Symmetric intensity diminutions such as these are often interpreted to result from shadows cast by a misaligned inner component of the circumstellar disc that intercepts light.
In Section \ref{sec:sed_fit} it is indeed established that a hot inner component is required to fit the SED, with a predicted location at  22$\pm$6~au. However, this predicted warm component is not sufficiently massive to provide the optical depth necessary to create such a large shadow. In order to be optically thick, the inner disc would have to be massive enough that thermal emission from the dust would be detectable with our observations. Furthermore, the continuum data is well fit by a symmetric model, as evidenced by the residual maps in Figure \ref{fig:tclean_images}. We cannot entirely rule out low-level underlying asymmetries in the brightness distribution that could be investigated with deeper observations, which if found could provide useful information for constraining the eccentricity of the ring or could signal interactions with an unseen planet.

We currently lack further resolved imaging that can help to characterise the system.
SPHERE observations of HD~126062 were made as part of the observing program presented in \citet{Matthews2018} but no detection of a debris disc was made. This is perhaps not surprising in light of the disc properties we derive here; total intensity observations are most successful at detecting edge-on discs where SNR in individual pixels reaches a maximum. Furthermore, ADI techniques will remove signal from a symmetric (or face-on) emission structure, and there is yet to be a debris disc detected with this method that has an inclination along the line of sight of the observer that is lower than 65$^\circ$ \citep{Matthews2023TheAssociation}. Three faint, wide-separation candidates are identified in the SPHERE images, but appear to be background objects \citep{Matthews2018}. Future observations using differential polarimetric imaging might offer a more promising route for detection of the disc in scattered light. These observations might detect pericenter or apocenter glow as seen in other systems. New observations could also clarify whether the intensity diminutions are due to shadowing from an inner disc component, as shadows would also appear in scattered light images, or whether the emission originates from a mostly symmetric ring of dust, the model which is consistent with current 1.3~mm ALMA observations as we demonstrate in this work.

\subsection{Constraints on planets in the system}

An unseen planet can clear gaps in the debris disc by scattering nearby particles throughout its orbit, and so for any well-defined debris disc density configuration we can infer what type of planets would have been able to create the observed distribution through gravitational interaction and migration. If the debris disc is relatively massive with respect to the planet, then the interchange of energy as a result of the disc-planet interactions will result in an inward migration motion for the planet \citep{Ida2000ORBITALOBJECTS,Pearce2014DynamicalDisc}. Alternatively, if the planet is relatively massive compared to the debris disc, then the dusty disc material will disperse quickly and will affect the planet's orbital semi-major axis to a minimal extent. \citet{Friebe2022GapDisc} explore the carving of debris disc gaps by planets in the context of these mass constraints and apply their results to a number of debris disc systems. They find that two solutions are found in each case; one of a high-mass planet that barely migrates, and another of a lower mass planet that migrates inward to a greater extent, clearing a gap in the debris disc as it does so. We adopt this approach of \citet{Friebe2022GapDisc} to apply constraints on the mass of an unseen planet residing in the HD~126062 system, assuming a single planet is responsible for the observed dust distribution. 

\begin{figure}
    \centering
    \includegraphics[width=1.0\linewidth]{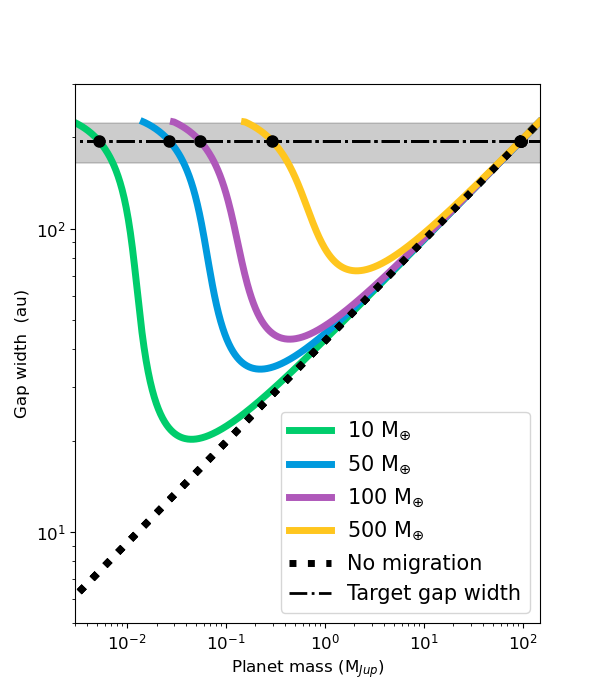}
    \caption{Gap widths opened by different masses of orbiting exoplanet, calculated for four different values of the initial debris disc mass. The horizontal line shows the target gap width, which has been measured from the dust distribution as defined by this work. The dotted line shows the result for a case in which there is no migration from the planet and the shaded area indicates the calculated uncertainty on the target gap width in the system. }
    \label{fig:planet_carve}
\end{figure}

For these calculations we describe the outer disc using the newly constrained exoKuiper belt parameters from Section \ref{sec:res_alma}, in addition to constraints on the inner, warm component of the debris disc that are derived from the SED fitting in Section \ref{sec:sed_fit}. Specifically, the target gap width is defined as the distance between the inner bound to the cold Kuiper belt and the outer limit placed on the inner disc emission by SED fitting. We calculate an uncertainty on this width by propagating the uncertainty on the width and position of the outer cool belt, and the uncertainty on the position of the inner disc. We therefore consider a target gap width of 195$\pm$58 au. The planetary orbits are assumed to be circular and in the same plane as the disc.
We consider an indicative range of initial debris disc masses at the point where the planet's influence begins to dominate the dust distribution in order to demonstrate the range of potential outcomes. Here we are informed by \citet{Krivov2021SolutionSmall}, who suggest 10 M$_\oplus$ as an approximate minimum mass for debris discs in order to successfully sustain a collisional cascade. The maximum dust mass of a debris disc is more difficult to constrain, and depends on a number of factors that are not yet well constrained, including the uncertain calculation of dust masses in protoplanetary discs, the rate of material replenishment in circumstellar discs and the fraction of total disc mass that is contained within solids. \citet{Krivov2021SolutionSmall} conclude that masses greater than $\approx$ 100-1000 M$_\oplus$ are less likely.
Figure \ref{fig:planet_carve} shows the results of these calculations, which give the size of the gap that is carved by a range of planet masses\footnote{These calculations are made using the general, or "exact" Equation 8 from \citet{Friebe2022GapDisc}, rather than the simplified version. The latter form nevertheless yields similar results and is valid for the majority of cases where migration is weak.}. The specific predicted planet masses that satisfy the dust distribution are summarised in Table \ref{tab:pot_planets}, in which two planetary masses are given, reflecting the degeneracy highlighted by \citet{Friebe2022GapDisc}. 

In addition to the results of Table \ref{tab:pot_planets} we also consider constraints placed by previous work on the system. There are upper limits from non-detections in SPHERE observations; \citet{Matthews2018} convert their contrast limits into upper limits on planet mass using COND \citep{Baraffe2003Evolutionary209458} and DUSTY \citep{Chabrier2000CoolingAtmospheres} evolutionary models, resulting in limits for direct imaging of $\approx$ 1 M$_{\rm Jup}$ at 10~au, or 2~M$_{\rm Jup}$ at 200~au. There are also limits from previous attempts to constrain putative planet masses through dynamices. For example HD~126062 is included in the work of \citet{Pearce2022PlanetSurveys}, where data from the planet-hunting ISPY, LEECH and LIStEN surveys are used in conjunction with descriptions of disc dust distribution from SED fitting which is similar to that performed in this work, except without our new, updated values for the radius and width of the outer belt. \citet{Pearce2022PlanetSurveys} estimate the minimum mass required of a single planet to carve the disc yet evade detection to be 7~M$_{\rm Jup}$, and also consider a multi-planet scenario, in which case they posit that each planet should be at least 3~M$_{\rm Jup}$. Both of these constraints are above the limits derived from the SPHERE non-detections. Our calculations using the distribution of mm brightness as constrained in this work suggest that planets responsible for sculpting the disc could be much less massive than this. For a massive initial disc the responsible planet could have a mass approaching that of Jupiter (e.g. top row Table \ref{tab:pot_planets}), whereas in the case for lower initial disc masses the predicted planet mass is more similar to Neptune- or Earth-mass planets. This demonstrates the importance of a well characterised outer disc when using our adopted method for inferring potential planet mass, which can be achieved through observations at mm/sub-mm wavelengths. Future observations with higher angular resolution will be able to constrain the distribution of material in the outer disc with even greater precision, which in turn would place tighter constraints on the mass of potential planets responsible for carving the observed dust distribution. Future observations would also improve the characterisation of the inner regions. 
Currently the uncertainty on the gap width is dominated by the error on the location of the inner disc, because it is relatively poorly constrained due to a lack of a spatially resolved detection. This could be addressed in the future via deep observations of the inner regions of the system, which might require higher frequency ALMA observations, or alternatively, mid-IR observations with JWST/MIRI.

The sensitivity limits of direct imaging \citep{Matthews2018} are consistent with the lower planet mass and stronger migration scenarios presented in Table \ref{tab:pot_planets}, all of which are below the detection limits and constraints set by previous work. The high mass, weaker migration scenario leads to very high planet masses, and can be ruled out when we consider the evidence from previous observations searching for exoplanets in the system. Therefore if a single, embedded exoplanet within the debris disc is responsible for carving a gap in the dust disc that we observe, the planet is likely to have had a relatively low mass (the first results column in Table  \ref{tab:pot_planets}) and migrated inwards to a relatively large extent. For example in the case of an initial disc mass of 100M$\oplus$, the planet would have to begin migrating and end at 212~au, and stop at around 33~au in order to explain the gap we observe using just a single planet. Our calculations here consider a single planet in order to place mass constraints. It may be that multiple lower mass planets exist in the HD~126062 system, and act in concert to carve the wider gap that has been observed. Future high sensitivity direct imaging campaigns would be required to confirm that planets reside in this system. 

In protoplanetary discs there is a correlation between measured disc masses and the spectral type of the host star \citep{Andrews2018ScalingDisks,Pascucci2016ARELATION}, intermediate mass T Tauris and Herbigs stars host more massive dust \citep{Stapper2022TheALMA} and indeed gas \citep{Stapper2024ConstrainingIsotopologues} discs than their lower mass counterparts. Greater masses in the protoplanetary phase provide a greater reservoir of solid particles which can eventually form part of the debris disc, as may have occurred around the A type star HD~126062, leading to the large exoKuiper belt that has been detected in the new ALMA observations. A high disc mass enables migration over larger distances, particularly if the planets are low in mass, as the results here favour for any putative planets around HD~126062.

Given that the system is relatively young (for a debris disc), any putative planet will likely be bright and thus a prime target for direct imaging. The constraints set in this work will aid with the design of future planet searches in HD~126062. Deeper planet searches in HD~126062 may yet provide a new arena in which the formation of giant planets can be investigated. This is particularly exciting given the detection of four wide orbit, giant planets around the very similar star of HR~8799, where ALMA observations also detected a large exoKuiper belt at a similar radius to that of HD~126062 (as shown in Figure \ref{fig:exoKuipers}).

By interpreting large exoKuiper belts as the result of interactions between the disc and planets on wide orbits, we enable the study of a population of planets that are the least likely to be detected by the most successful modes of planet detection such as the radial velocity and transit methods. In this way debris discs are a key resource in studying the formation of planets with relatively long orbits. In the case of HD~126062, if the distribution of solids in the large debris disc has resulted from the shepherding and stirring of a single planet, our results suggest that planet must have formed in the outer regions of the young disc, and migrated inward. This might initially suggest that the putative planet is unlikely to have formed via the widely favoured core-accretion model of planet formation, under which the timescales for the accumulation of disc material in the outer regions are simply too long. However, formation via the gravitational instability method is predicted to produce giant planets \citep{Kratter2010TheStars,Forgan2011TheMass}, whereas the constraints set here prefer sub-Jupiter masses. Even in the case of planetary system HR~8799, in which all the detected planets are super-Jovian in mass, \citet{Kratter2010TheStars} point out that a number of stringent criteria must be met for GI to remain the favoured formation pathway. To disentangle these competing scenarios, further observational constraints are essential. In particular, deeper imaging campaigns targeting wide-orbit planets, together with systematic surveys of debris discs around A-type stars, will be crucial for testing whether gravitational instability or core accretion is the dominant mechanism shaping these planetary systems.

\begin{table}
\caption{The specific masses calculated for planets capable of carving the HD~126062 dust distribution through their orbital motion, assuming a circular orbit and a range of potential initial disc masses.  }
\centering\begin{tabular}{|c|c|c|}
\hline
\begin{tabular}[c]{@{}c@{}}\textbf{Initial Disc} \\ \textbf{Mass} \\ (M$_\oplus$)\end{tabular} & \begin{tabular}[c]{@{}c@{}}\textbf{Low mass,} \\ \textbf{strong migration}\\ (M$_{\rm Jup}$)\end{tabular} & \begin{tabular}[c]{@{}c@{}}\textbf{High mass,} \\ \textbf{weak migration}\\ (M$_{\rm Jup}$)\end{tabular} \\ \hline
10 & 0.005 & 96.2 \\
50 & 0.027 & 96.2 \\
100 & 0.056 & 96.1 \\
500 & 0.293 & 94.6 \\ \hline
\end{tabular}

\label{tab:pot_planets}
\end{table}

\subsection{\textbf{$^{12}$CO (2--1) emission in the vicinity of HD~126062}}
\label{sec:discussionGas}

\subsubsection{A lack of evidence connecting gas emission to HD~126062}

We now consider potential interpretations of the $^{12}$CO(2--1) emission that is detected and assess to what extent they are consistent with the data at hand on HD~126062. The observed properties of the emission are not consistent with the typical scenarios invoked to explain gas in debris discs, as we explore in the following section. 

The regions of peak brightness in the $^{12}$CO(2--1) emission are detected at greater separation from the star than the exo-Kuiper belt, and are broadly aligned with the two faintest parts of the continuum ring. This naturally invites speculation of a physical connection between the two, but as was discussed in Section \ref{sec:res_alma}, results from visibility fitting of the continuum data strongly favour an inclination closer to face-on (<17$^\circ$). With current data we therefore favour scenarios in which, in terms of projection on the sky, the gas is mainly external to a face-on debris disc, rather than one in which the emitting gas is co-spatial with a disc that is more extended than the continuum belt.

Very little emission is detected from regions that are interior to the exo-Kuiper belt, whereas significant $^{12}$CO(2--1) intensity is mostly detected externally to the disc. 
Models of gas in debris discs do not predict gas that is found only in locations radially external to the dust distribution, as is the case in the images in Figures \ref{fig:mom0} and \ref{fig:spectra_and_mom8}. In some systems the gas emission is observed to extend further inwards than the detectable levels of mm dust emission \citep[e.g. HD~21997, 49 Ceti,][]{Kospal2013, Moor2013, Hughes2017RadialCeti}, but here we find the reverse situation. 
If the gas in the debris disc is secondary, one would generally expect the gas and dust to be largely co-spatial, as it is from collisions of planetesimals within the dust belts that gas is released. Although some viscous spreading does occur from this birth-site within the belt \citep{Marino2020PopulationStars}. In a primordial gas scenario, the configuration of gas and dust could be more similar to that seen in protoplanetary discs, where the gas disc is much more extended than the dust disc \citep[e.g.][]{Trapman2019GasEvolution}. However, in protoplanetary discs the gas density profile is usually centrally peaked, whereas in Figure \ref{fig:mom0} there is a lack of emission from within the inner regions. The lack of emission from within the ring of HD~126062 is puzzling, as we would ordinarily expect the regions closest to the star to be the hottest, and more likely to be emitting line emission, rather than those at greater radial separation, and so such a scenario seems difficult to achieve. It could be that the CO actually originates from an object in the background to HD~126062 and that emission is being blocked along the line of sight of the observer in order to result in an apparent inner cavity of gas emission. This might occur if the disc contained large quantities of CO that absorbs the CO radiation from the background, but is so cold that it does not emit strongly itself.

Perhaps then, the gas we detect has recently exited the HD~126062 system. It is unlikely in this case that photoevaporative dispersal is at play. Given the spectral type of the star and the distance to the exoKuiper belt, the gas could not be heated to a temperature that would unbind it gravitationally from the host star; at the position of the exoKuiper belt, R=270~au, this would require a temperature $\sim$3139~K. Furthermore we would expect a much greater velocity displacement from that of the system in the case of a wind.
Alternatively, gas may leave the system via a wind. \citet{Kral2023PotentialWinds} model the potential effects of stellar winds on gas residing in the belts of debris discs \citep[as is predicted to occur in our own Kuiper belt,][]{Kral2021ABelt} and identify that winds are expected in discs with L$_* > 20 L_\odot$, in which case the gas will form a wind rather than a disc. Stellar winds may also occur in systems with a luminosity below this criterion in low density environments \citep[$n_{\rm crit} \approx 10^{-3}$cm$^{-3}$,][]{Kral2023PotentialWinds}. The latter scenario would apply in the case of HD~126062; our SED fit constrains the stellar luminosity to 15.5~L$_\odot$. 
The spectrum of an emission line tracing such a wind is expected to be broad however, for example as was detected in the Class III system NO Lup \citep{Lovell2021RapidDisc}, or further offset with respect to the system velocity \cite[e.g. the wind in $\eta$ Tel, with components blue shifted by $\approx17-23~ \rm km/s$ shown by ][]{Youngblood2021ADisk}. We note however a dependence on inclination of the source; whilst $\eta$ Tel is highly inclined, HD~126062 is close to face on, and so a wind cannot be entirely ruled out based on velocity alone. Kinematic analysis of gas emission can indicate relative motions of the gas, which can be compared to the predictions from models of belt winds. In the case of HD~126062 this may be difficult; radiatively driven winds have a velocity vector in the radial direction, which for a nearly face-on disc will be projected as almost perpendicular to the line of sight of the observer, minimising any observable Doppler shift. The CO spectrum of HD~126062 is very narrow, with a line width of $\approx$2~km/s compared to that of NO Lup $\approx$20km/s \citep{Lovell2021RapidDisc}, and so future follow-up observations should leverage the finest possible spectral resolution in order to resolve the gas emission in the vicinity of HD~126062, so that further investigation can be undertaken.

The moment 8 map in Figure \ref{fig:spectra_and_mom8} is somewhat reminiscent of double-spirals that can appear in the aftermath of a flyby interaction \citep[see e.g. ][]{Cuello2020FlybysSignatures}, however we lack a culprit companion with which HD~126062 could have interacted. The closest sources to HD~126062 on the sky are two stars at $\approx 7''$, but the proper motions and parallaxes of these two stars are inconsistent with membership of the association \citet{Luhman2022AComplex}. Additionally the potential companion candidates identified in SPHERE data were also classified as background objects \citep{Matthews2018}.

\subsubsection{Alternative explanations}

Given the lack of clear, `smoking-gun' evidence that ties the gas to the system of HD~126062, we must also consider the possibility that the gas emission is not in fact associated with HD~126062 at all. A similar situation was identified in observations of HR~8799, whereby $^{12}$CO  emission was detected by ALMA within the field of view, but was attributed to background cloud HLCG 92-35 \citep{Faramaz2021A7}.

The spectrum of the $^{12}$CO emission is shifted slightly in the negative direction, with the peak occurring at -1~km/s LSRK. But as shown in Figure \ref{fig:spectra_and_mom8} the systemic velocity of the source is at 7.2 km/s. If the gas emission was associated with the debris disc it would be expected to possess the same velocity as the star. This displacement in velocity space is more consistent with a scenario in which the gas emission is not associated with HD~126062. The fact that the line is resolved across three channels makes it unlikely that the feature results from incorrect telluric correction, as this is typically a much narrower feature ($<$ 30 kHz) which would have no velocity shift and would appear ubiquitous and extended in the image. Having inspected the Tsys and bandpass calibration, there is no evidence of a telluric line or of any calibration artifact. Furthermore, other targets observed within the same scheduling block do not show a similar feature and the line is detected in both of the individual observations, which makes it very unlikely that the emission is a bandpass artifact.
It could be that the emission originates from a background cloud.
If the velocity offset from the rest frequency of $^{12}$CO (2--1) was a result of movement due to Galactic rotation, we would be able to pinpoint the distance to this emitting cloud. However, the small shift of the line spectrum of -1~km/s from the rest frequency of $^{12}$CO is not consistent with anything that is predicted when considering the Galaxy's rotation curve as measured in the literature \citep[for example,][]{Reid2019TrigonometricWay, Eilers2019TheKpc}. Cloud-to-cloud dispersion velocities are often measured as being of order a few km/s \citep{Wilson2011TheGalaxies,Sun2020MolecularPopulation,Spilker2022BirdsScales}, and so a relatively low offset such as -1~km/s could easily result from peculiar motion in such a cloud as viewed along the line of sight of the observer, but in this case we cannot know its relative distance from Earth. 
There is also the possibility that the emission could originate from a foreground or background object, but the lack of emission in the centre of the image where the debris disc resides, makes it difficult to imagine a foreground object that could produce such a morphology, other than diffuse clouds either in the vicinity of the star or along the line of sight.

Both the continuum emission from the debris disc and the extended $^{12}$CO emission are centred on the stellar position, but this does not necessitate that the two emission structures originate from the same object \citep[as was seen towards HR~8799 by][]{Faramaz2021A7}. Whilst the origin and nature of the extended gas emission remains unclear with current data, targeted follow-up observations will be able to distinguish between, or rule-out some of the possibilities discussed in this section. High spectral resolution observations would enable investigation into the kinematic motion of the gas which could shed light on the nature of the emitting object. Higher angular resolution observations would provide a more precise mapping of the morphology of the brightness distribution in the emitting object, which would help to identify it.

\section{Conclusions}

In this article we present ALMA continuum imaging that has detected the exoKuiper belt around HD~126062 at mm wavelengths for the first time. Using Monte Carlo sampling we make fits of a Gaussian ring morphology to the observed visibilities, finding a relatively face-on configuration with inclination of $\leq 17^\circ$. The observed visibilities are consistent with a model of a Gaussian ring with a radial separation of the ring from the stellar position of 2.01$^{+0.04}_{-0.05}$$''$ and a standard deviation of 0.30$^{+0.04}_{-0.04}$$''$. Considering the distance to HD~126062 and converting the standard deviation of the Gaussian ring to a full width at half maximum, the exoKuiper parameters can be expressed in spatial units with radius $R=270^{+5}_{-4}$~au and width $\Delta R=95\pm12$~au. This is one of the largest known exoKuiper belts to be resolved at millimetre wavelengths to date.

In addition to the exoKuiper belt, a second, warm component to the HD~126062 debris disc is required to achieve an accurate fit to the SED of the object, which we update with the new 1.3~mm photometric point. Our two-temperature blackbody fit predicts this inner component to be found at 22$\pm$6~au.

We detect extended $^{12}$CO emission in the vicinity of the star and debris disc, where the majority of gas line mission is situated externally to the exoKuiper belt we have detected. The exact nature of the $^{12}$CO emission remains unclear, it seems most likely that the gas is not associated with HD~126062, future observations are required to confirm the current working hypothesis. 

Using the newly constrained dust distribution in the HD~126062 debris disc, we put limits on the mass of a potential unseen planet that could carve such a configuration through their orbital motion. Considering the constraints set by previous, unsuccessful planet searches, the dust distribution we observed in HD~126062 is consistent with sculpting by planets with a sub-Jovian planet ($\lesssim$ M$_{\rm J}$) that has migrated significantly through the disc. Alternatively a number of smaller planets could exist within the gap in dust distribution. 

Future observations of this object are strongly encouraged in order to more closely characterise the distribution of dust in the debris disc, and to further study the $^{12}$CO gas that appears near to HD~126062 in the images presented here. Specifically, deeper mm or sub-mm observations with high angular resolution should seek to precisely define the width and eccentricity of the exoKuiper belt, detect the predicted inner component of the debris disc, and to further investigate the true nature of the $^{12}$CO emission in the vicinity of HD~126062.

\begin{acknowledgements}

This paper makes use of the following ALMA data: ADS/JAO.ALMA$\#$2022.1.00968.S. ALMA is a partnership of ESO (representing its member states), NSF (USA) and NINS (Japan), together with NRC (Canada), NSTC and ASIAA (Taiwan), and KASI (Republic of Korea), in cooperation with the Republic of Chile. The Joint ALMA Observatory is operated by ESO, AUI/NRAO and NAOJ. This research made use of numpy \citep{Harris2020ArrayNumPy} and Astropy,\footnote{http://www.astropy.org} a community-developed core Python package for Astronomy \citep{TheAstropyCollaboration2013,TheAstropyCollaboration2018ThePackage} and the SIMBAD database, operated at CDS, Strasbourg, France. The authors would additionally like to thank Elisabeth Matthews for discussions on the SPHERE observations of HD~126062 and Seiji Kameno for his insight on telluric features in ALMA data. The authors acknowledge support from ANID -- Millennium Science Initiative Program -- Center Code NCN2024\_001. J.M. acknowledges support from FONDECYT de Postdoctorado 2024 \#3240612. A.R. has been supported by the UK Science and Technology Facilities Council (STFC) via the consolidated grant ST/W000997/1 and by the European Union’s Horizon 2020 research and innovation programme under the Marie Sklodowska-Curie grant agreement No. 823823 (RISE DUSTBUSTERS project). TJH acknowledges UKRI guaranteed funding for a Horizon Europe ERC consolidator grant (EP/Y024710/1) and a Royal Society Dorothy Hodgkin Fellowship. J.M. additionally wishes to thank Instituto Roca Negra for their hospitality and fruitful discussions during the conception of this investigation.

\end{acknowledgements}

\bibliography{aa54463-25.bib}

@article{Luhman2022AComplex,
    title = {{A Census of the Stellar Populations in the Sco-Cen Complex*}},
    year = {2022},
    journal = {AJ},
    author = {Luhman, K. L.},
    number = {1},
    month = {1},
    pages = {24},
    volume = {163},
    publisher = {IOP Publishing},
    url = {https://iopscience.iop.org/article/10.3847/1538-3881/ac35e2},
    doi = {10.3847/1538-3881/ac35e2},
    issn = {0004-6256}
}

@article{MacGregor2017,
    title = {{A Complete ALMA Map of the Fomalhaut Debris Disk}},
    year = {2017},
    journal = {ApJ},
    author = {MacGregor, Meredith A. and Matr{\`{a}}, Luca and Kalas, Paul and Wilner, David J. and Pan, Margaret and Kennedy, Grant M. and Wyatt, Mark C. and Duchene, Gaspard and Hughes, A. Meredith and Rieke, George H. and Clampin, Mark and Fitzgerald, Michael P. and Graham, James R. and Holland, Wayne S. and Pani{\'{c}}, Olja and Shannon, Andrew and Su, Kate},
    number = {1},
    month = {6},
    pages = {8},
    volume = {842},
    url = {https://iopscience.iop.org/article/10.3847/1538-4357/aa71ae},
    doi = {10.3847/1538-4357/aa71ae},
    issn = {0004-637X}
}

@article{Kral2021ABelt,
    title = {{A molecular wind blows out of the Kuiper belt}},
    year = {2021},
    journal = {A{\&}A},
    author = {Kral, Quentin and Pringle, J. E. and Guilbert-Lepoutre, Aurélie and Matr{\`{a}}, Luca and Moses, Julianne I. and Lellouch, Emmanuel and Wyatt, Mark C. and Biver, Nicolas and Bockel{\'{e}}e-Morvan, Dominique and Bonsor, Amy and Petit, Franck Le and Gladstone, G. Randall},
    month = {4},
    pages = {L11},
    volume = {653},
    url = {http://arxiv.org/abs/2104.01035 http://dx.doi.org/10.1051/0004-6361/202141783},
    doi = {10.1051/0004-6361/202141783},
    arxivId = {2104.01035}
}

@article{Lagrange2009AImaging,
    title = {{A probable giant planet imaged in the {\{}{\$}{$\beta$}{\$}{\}} Pictoris disk* VLT/NaCo deep L′-band imaging}},
    year = {2009},
    journal = {A\&A},
    author = {Lagrange, A M and Gratadour, D and Chauvin, G and Fusco, T and Ehrenreich, D and Mouillet, D and Rousset, G and Rouan, D and Allard, F and Gendron, É and Charton, J and Mugnier, L and Rabou, P and Montri, J and Lacombe, F},
    number = {2},
    pages = {21--25},
    volume = {493},
    doi = {10.1051/0004-6361:200811325},
    issn = {00046361}
}

@article{Youngblood2021ADisk,
    title = {{A Radiatively Driven Wind from the eta Tel Debris Disk}},
    year = {2021},
    journal = {AJ},
    author = {Youngblood, Allison and Roberge, Aki and MacGregor, Meredith A. and Brandeker, Alexis and Weinberger, Alycia and P{\'{e}}rez, Sebastián and Grady, Carol and Welsh, Barry},
    number = {6},
    month = {8},
    pages = {235},
    volume = {162},
    url = {http://arxiv.org/abs/2108.11965 http://dx.doi.org/10.3847/1538-3881/ac21d1},
    doi = {10.3847/1538-3881/ac21d1},
    arxivId = {2108.11965}
}

@article{Moor2013,
    title = {{A RESOLVED DEBRIS DISK AROUND THE CANDIDATE PLANET-HOSTING STAR HD 95086}},
    year = {2013},
    journal = {ApJ},
    author = {Mo{\'{o}}r, A. and Abrah{\'{a}}m, P. and K{\'{o}}sp{\'{a}}l, A. and Szabo, Gy. M. and Apai, D. and Balog, Z. and Csengeri, T. and Grady, C. and Henning, Th. and Juh{\'{a}}sz, A. and Kiss, Cs. and Pascucci, I. and Szul{\'{a}}gyi, J. and Vavrek, R.},
    number = {2},
    month = {9},
    pages = {L51},
    volume = {775},
    publisher = {IOP Publishing},
    url = {http://stacks.iop.org/2041-8205/775/i=2/a=L51?key=crossref.f108317d5d66d9c54024c279fffc8fbb},
    doi = {10.1088/2041-8205/775/2/L51}
}

@article{White2016,
    title = {{ALMA OBSERVATIONS OF HD 141569’s CIRCUMSTELLAR DISK}},
    year = {2016},
    journal = {ApJ},
    author = {White, J. A. and Boley, A. C. and Hughes, A. M. and Flaherty, K. M. and Ford, E. and Wilner, D. and Corder, S. and Payne, M.},
    number = {1},
    month = {9},
    pages = {6},
    volume = {829},
    publisher = {IOP Publishing},
    url = {http://stacks.iop.org/0004-637X/829/i=1/a=6?key=crossref.0ba434c9061c3c33a96972222c247943},
    doi = {10.3847/0004-637X/829/1/6},
    issn = {1538-4357}
}

@article{Kospal2013,
    title = {{ALMA OBSERVATIONS OF THE MOLECULAR GAS IN THE DEBRIS DISK OF THE 30 Myr OLD STAR HD 21997}},
    year = {2013},
    journal = {ApJ},
    author = {Kospal, A and Moor, A. and Juhasz, A. and Abraham, P. and Apai, D. and Csengeri, T. and Grady, C. A. and Henning, Th. and Hughes, A. M. and Kiss, Cs. and Pascucci, I. and Schmalzl, M.},
    number = {2},
    month = {9},
    pages = {77},
    volume = {776},
    publisher = {IOP Publishing},
    url = {http://stacks.iop.org/0004-637X/776/i=2/a=77?key=crossref.8f3d110723821b6af2bbb836cc8aba45},
    doi = {10.1088/0004-637X/776/2/77},
    issn = {0004-637X}
}

@article{Vallenari2023iGaia/i3,
    title = {{<i>Gaia</i> Data Release 3}},
    year = {2023},
    journal = {A{\&}A},
    author = {Vallenari, A. and Brown, A. G. A. and Prusti, T. and de Bruijne, J. H. J. and Arenou, F. and Babusiaux, C. and Biermann, M. and Creevey, O. L. and Ducourant, C. and Evans, D. W. and Eyer, L. and Guerra, R. and Hutton, A. and Jordi, C. and Klioner, S. A. and Lammers, U. L. and Lindegren, L. and Luri, X. and Mignard, F. and Panem, C. and Pourbaix, D. and Randich, S. and Sartoretti, P. and Soubiran, C. and Tanga, P. and Walton, N. A. and Bailer-Jones, C. A. L. and Bastian, U. and Drimmel, R. and Jansen, F. and Katz, D. and Lattanzi, M. G. and van Leeuwen, F. and Bakker, J. and Cacciari, C. and Casta{\~{n}}eda, J. and De Angeli, F. and Fabricius, C. and Fouesneau, M. and Fr{\'{e}}mat, Y. and Galluccio, L. and Guerrier, A. and Heiter, U. and Masana, E. and Messineo, R. and Mowlavi, N. and Nicolas, C. and Nienartowicz, K. and Pailler, F. and Panuzzo, P. and Riclet, F. and Roux, W. and Seabroke, G. M. and Sordo, R. and Th{\'{e}}venin, F. and Gracia-Abril, G. and Portell, J. and Teyssier, D. and Altmann, M. and Andrae, R. and Audard, M. and Bellas-Velidis, I. and Benson, K. and Berthier, J. and Blomme, R. and Burgess, P. W. and Busonero, D. and Busso, G. and C{\'{a}}novas, H. and Carry, B. and Cellino, A. and Cheek, N. and Clementini, G. and Damerdji, Y. and Davidson, M. and de Teodoro, P. and Nu{\~{n}}ez Campos, M. and Delchambre, L. and Dell’Oro, A. and Esquej, P. and Fern{\'{a}}ndez-Hern{\'{a}}ndez, J. and Fraile, E. and Garabato, D. and Garc{\'{i}}a-Lario, P. and Gosset, E. and Haigron, R. and Halbwachs, J.-L. and Hambly, N. C. and Harrison, D. L. and Hern{\'{a}}ndez, J. and Hestroffer, D. and Hodgkin, S. T. and Holl, B. and Jan{\ss}en, K. and Jevardat de Fombelle, G. and Jordan, S. and Krone-Martins, A. and Lanzafame, A. C. and L{\"{o}}ffler, W. and Marchal, O. and Marrese, P. M. and Moitinho, A. and Muinonen, K. and Osborne, P. and Pancino, E. and Pauwels, T. and Recio-Blanco, A. and Reyl{\'{e}}, C. and Riello, M. and Rimoldini, L. and Roegiers, T. and Rybizki, J. and Sarro, L. M. and Siopis, C. and Smith, M. and Sozzetti, A. and Utrilla, E. and van Leeuwen, M. and Abbas, U. and {\'{A}}brah{\'{a}}m, P. and Abreu Aramburu, A. and Aerts, C. and Aguado, J. J. and Ajaj, M. and Aldea-Montero, F. and Altavilla, G. and {\'{A}}lvarez, M. A. and Alves, J. and Anders, F. and Anderson, R. I. and Anglada Varela, E. and Antoja, T. and Baines, D. and Baker, S. G. and Balaguer-N{\'{u}}{\~{n}}ez, L. and Balbinot, E. and Balog, Z. and Barache, C. and Barbato, D. and Barros, M. and Barstow, M. A. and Bartolom{\'{e}}, S. and Bassilana, J.-L. and Bauchet, N. and Becciani, U. and Bellazzini, M. and Berihuete, A. and Bernet, M. and Bertone, S. and Bianchi, L. and Binnenfeld, A. and Blanco-Cuaresma, S. and Blazere, A. and Boch, T. and Bombrun, A. and Bossini, D. and Bouquillon, S. and Bragaglia, A. and Bramante, L. and Breedt, E. and Bressan, A. and Brouillet, N. and Brugaletta, E. and Bucciarelli, B. and Burlacu, A. and Butkevich, A. G. and Buzzi, R. and Caffau, E. and Cancelliere, R. and Cantat-Gaudin, T. and Carballo, R. and Carlucci, T. and Carnerero, M. I. and Carrasco, J. M. and Casamiquela, L. and Castellani, M. and Castro-Ginard, A. and Chaoul, L. and Charlot, P. and Chemin, L. and Chiaramida, V. and Chiavassa, A. and Chornay, N. and Comoretto, G. and Contursi, G. and Cooper, W. J. and Cornez, T. and Cowell, S. and Crifo, F. and Cropper, M. and Crosta, M. and Crowley, C. and Dafonte, C. and Dapergolas, A. and David, M. and David, P. and de Laverny, P. and De Luise, F. and De March, R. and De Ridder, J. and de Souza, R. and de Torres, A. and del Peloso, E. F. and del Pozo, E. and Delbo, M. and Delgado, A. and Delisle, J.-B. and Demouchy, C. and Dharmawardena, T. E. and Di Matteo, P. and Diakite, S. and Diener, C. and Distefano, E. and Dolding, C. and Edvardsson, B. and Enke, H. and Fabre, C. and Fabrizio, M. and Faigler, S. and Fedorets, G. and Fernique, P. and Fienga, A. and Figueras, F. and Fournier, Y. and Fouron, C. and Fragkoudi, F. and Gai, M. and Garcia-Gutierrez, A. and Garcia-Reinaldos, M. and Garc{\'{i}}a-Torres, M. and Garofalo, A. and Gavel, A. and Gavras, P. and Gerlach, E. and Geyer, R. and Giacobbe, P. and Gilmore, G. and Girona, S. and Giuffrida, G. and Gomel, R. and Gomez, A. and Gonz{\'{a}}lez-N{\'{u}}{\~{n}}ez, J. and Gonz{\'{a}}lez-Santamar{\'{i}}a, I. and Gonz{\'{a}}lez-Vidal, J. J. and Granvik, M. and Guillout, P. and Guiraud, J. and Guti{\'{e}}rrez-S{\'{a}}nchez, R. and Guy, L. P. and Hatzidimitriou, D. and Hauser, M. and Haywood, M. and Helmer, A. and Helmi, A. and Sarmiento, M. H. and Hidalgo, S. L. and Hilger, T. and H{\l}adczuk, N. and Hobbs, D. and Holland, G. and Huckle, H. E. and Jardine, K. and Jasniewicz, G. and Jean-Antoine Piccolo, A. and Jim{\'{e}}nez-Arranz, Ó. and Jorissen, A. and Juaristi Campillo, J. and Julbe, F. and Karbevska, L. and Kervella, P. and Khanna, S. and Kontizas, M. and Kordopatis, G. and Korn, A. J. and K{\'{o}}sp{\'{a}}l, Á and Kostrzewa-Rutkowska, Z. and Kruszy{\'{n}}ska, K. and Kun, M. and Laizeau, P. and Lambert, S. and Lanza, A. F. and Lasne, Y. and Le Campion, J.-F. and Lebreton, Y. and Lebzelter, T. and Leccia, S. and Leclerc, N. and Lecoeur-Taibi, I. and Liao, S. and Licata, E. L. and Lindstr{\o}m, H. E. P. and Lister, T. A. and Livanou, E. and Lobel, A. and Lorca, A. and Loup, C. and Madrero Pardo, P. and Magdaleno Romeo, A. and Managau, S. and Mann, R. G. and Manteiga, M. and Marchant, J. M. and Marconi, M. and Marcos, J. and Marcos Santos, M. M. S. and Mar{\'{i}}n Pina, D. and Marinoni, S. and Marocco, F. and Marshall, D. J. and Martin Polo, L. and Mart{\'{i}}n-Fleitas, J. M. and Marton, G. and Mary, N. and Masip, A. and Massari, D. and Mastrobuono-Battisti, A. and Mazeh, T. and McMillan, P. J. and Messina, S. and Michalik, D. and Millar, N. R. and Mints, A. and Molina, D. and Molinaro, R. and Moln{\'{a}}r, L. and Monari, G. and Mongui{\'{o}}, M. and Montegriffo, P. and Montero, A. and Mor, R. and Mora, A. and Morbidelli, R. and Morel, T. and Morris, D. and Muraveva, T. and Murphy, C. P. and Musella, I. and Nagy, Z. and Noval, L. and Oca{\~{n}}a, F. and Ogden, A. and Ordenovic, C. and Osinde, J. O. and Pagani, C. and Pagano, I. and Palaversa, L. and Palicio, P. A. and Pallas-Quintela, L. and Panahi, A. and Payne-Wardenaar, S. and Pe{\~{n}}alosa Esteller, X. and Penttil{\"{a}}, A. and Pichon, B. and Piersimoni, A. M. and Pineau, F.-X. and Plachy, E. and Plum, G. and Poggio, E. and Pr{\v{s}}a, A. and Pulone, L. and Racero, E. and Ragaini, S. and Rainer, M. and Raiteri, C. M. and Rambaux, N. and Ramos, P. and Ramos-Lerate, M. and Re Fiorentin, P. and Regibo, S. and Richards, P. J. and Rios Diaz, C. and Ripepi, V. and Riva, A. and Rix, H.-W. and Rixon, G. and Robichon, N. and Robin, A. C. and Robin, C. and Roelens, M. and Rogues, H. R. O. and Rohrbasser, L. and Romero-G{\'{o}}mez, M. and Rowell, N. and Royer, F. and Ruz Mieres, D. and Rybicki, K. A. and Sadowski, G. and S{\'{a}}ez N{\'{u}}{\~{n}}ez, A. and Sagrist{\`{a}} Sell{\'{e}}s, A. and Sahlmann, J. and Salguero, E. and Samaras, N. and Sanchez Gimenez, V. and Sanna, N. and Santove{\~{n}}a, R. and Sarasso, M. and Schultheis, M. and Sciacca, E. and Segol, M. and Segovia, J. C. and S{\'{e}}gransan, D. and Semeux, D. and Shahaf, S. and Siddiqui, H. I. and Siebert, A. and Siltala, L. and Silvelo, A. and Slezak, E. and Slezak, I. and Smart, R. L. and Snaith, O. N. and Solano, E. and Solitro, F. and Souami, D. and Souchay, J. and Spagna, A. and Spina, L. and Spoto, F. and Steele, I. A. and Steidelm{\"{u}}ller, H. and Stephenson, C. A. and S{\"{u}}veges, M. and Surdej, J. and Szabados, L. and Szegedi-Elek, E. and Taris, F. and Taylor, M. B. and Teixeira, R. and Tolomei, L. and Tonello, N. and Torra, F. and Torra, J. and Torralba Elipe, G. and Trabucchi, M. and Tsounis, A. T. and Turon, C. and Ulla, A. and Unger, N. and Vaillant, M. V. and van Dillen, E. and van Reeven, W. and Vanel, O. and Vecchiato, A. and Viala, Y. and Vicente, D. and Voutsinas, S. and Weiler, M. and Wevers, T. and Wyrzykowski, Ł. and Yoldas, A. and Yvard, P. and Zhao, H. and Zorec, J. and Zucker, S. and Zwitter, T.},
    month = {6},
    pages = {A1},
    volume = {674},
    publisher = {EDP Sciences},
    url = {https://www.aanda.org/10.1051/0004-6361/202243940},
    doi = {10.1051/0004-6361/202243940},
    issn = {0004-6361}
}

@article{Brown2021iGaia/i3,
    title = {{<i>Gaia</i> Early Data Release 3}},
    year = {2021},
    journal = {A{\&}A},
    author = {Brown, A. G. A. and Vallenari, A. and Prusti, T. and de Bruijne, J. H. J. and Babusiaux, C. and Biermann, M. and Creevey, O. L. and Evans, D. W. and Eyer, L. and Hutton, A. and Jansen, F. and Jordi, C. and Klioner, S. A. and Lammers, U. and Lindegren, L. and Luri, X. and Mignard, F. and Panem, C. and Pourbaix, D. and Randich, S. and Sartoretti, P. and Soubiran, C. and Walton, N. A. and Arenou, F. and Bailer-Jones, C. A. L. and Bastian, U. and Cropper, M. and Drimmel, R. and Katz, D. and Lattanzi, M. G. and van Leeuwen, F. and Bakker, J. and Cacciari, C. and Casta{\~{n}}eda, J. and De Angeli, F. and Ducourant, C. and Fabricius, C. and Fouesneau, M. and Fr{\'{e}}mat, Y. and Guerra, R. and Guerrier, A. and Guiraud, J. and Jean-Antoine Piccolo, A. and Masana, E. and Messineo, R. and Mowlavi, N. and Nicolas, C. and Nienartowicz, K. and Pailler, F. and Panuzzo, P. and Riclet, F. and Roux, W. and Seabroke, G. M. and Sordo, R. and Tanga, P. and Th{\'{e}}venin, F. and Gracia-Abril, G. and Portell, J. and Teyssier, D. and Altmann, M. and Andrae, R. and Bellas-Velidis, I. and Benson, K. and Berthier, J. and Blomme, R. and Brugaletta, E. and Burgess, P. W. and Busso, G. and Carry, B. and Cellino, A. and Cheek, N. and Clementini, G. and Damerdji, Y. and Davidson, M. and Delchambre, L. and Dell’Oro, A. and Fern{\'{a}}ndez-Hern{\'{a}}ndez, J. and Galluccio, L. and Garc{\'{i}}a-Lario, P. and Garcia-Reinaldos, M. and Gonz{\'{a}}lez-N{\'{u}}{\~{n}}ez, J. and Gosset, E. and Haigron, R. and Halbwachs, J.-L. and Hambly, N. C. and Harrison, D. L. and Hatzidimitriou, D. and Heiter, U. and Hern{\'{a}}ndez, J. and Hestroffer, D. and Hodgkin, S. T. and Holl, B. and Jan{\ss}en, K. and Jevardat de Fombelle, G. and Jordan, S. and Krone-Martins, A. and Lanzafame, A. C. and L{\"{o}}ffler, W. and Lorca, A. and Manteiga, M. and Marchal, O. and Marrese, P. M. and Moitinho, A. and Mora, A. and Muinonen, K. and Osborne, P. and Pancino, E. and Pauwels, T. and Petit, J.-M. and Recio-Blanco, A. and Richards, P. J. and Riello, M. and Rimoldini, L. and Robin, A. C. and Roegiers, T. and Rybizki, J. and Sarro, L. M. and Siopis, C. and Smith, M. and Sozzetti, A. and Ulla, A. and Utrilla, E. and van Leeuwen, M. and van Reeven, W. and Abbas, U. and Abreu Aramburu, A. and Accart, S. and Aerts, C. and Aguado, J. J. and Ajaj, M. and Altavilla, G. and {\'{A}}lvarez, M. A. and {\'{A}}lvarez Cid-Fuentes, J. and Alves, J. and Anderson, R. I. and Anglada Varela, E. and Antoja, T. and Audard, M. and Baines, D. and Baker, S. G. and Balaguer-N{\'{u}}{\~{n}}ez, L. and Balbinot, E. and Balog, Z. and Barache, C. and Barbato, D. and Barros, M. and Barstow, M. A. and Bartolom{\'{e}}, S. and Bassilana, J.-L. and Bauchet, N. and Baudesson-Stella, A. and Becciani, U. and Bellazzini, M. and Bernet, M. and Bertone, S. and Bianchi, L. and Blanco-Cuaresma, S. and Boch, T. and Bombrun, A. and Bossini, D. and Bouquillon, S. and Bragaglia, A. and Bramante, L. and Breedt, E. and Bressan, A. and Brouillet, N. and Bucciarelli, B. and Burlacu, A. and Busonero, D. and Butkevich, A. G. and Buzzi, R. and Caffau, E. and Cancelliere, R. and C{\'{a}}novas, H. and Cantat-Gaudin, T. and Carballo, R. and Carlucci, T. and Carnerero, M. I and Carrasco, J. M. and Casamiquela, L. and Castellani, M. and Castro-Ginard, A. and Castro Sampol, P. and Chaoul, L. and Charlot, P. and Chemin, L. and Chiavassa, A. and Cioni, M.-R. L. and Comoretto, G. and Cooper, W. J. and Cornez, T. and Cowell, S. and Crifo, F. and Crosta, M. and Crowley, C. and Dafonte, C. and Dapergolas, A. and David, M. and David, P. and de Laverny, P. and De Luise, F. and De March, R. and De Ridder, J. and de Souza, R. and de Teodoro, P. and de Torres, A. and del Peloso, E. F. and del Pozo, E. and Delbo, M. and Delgado, A. and Delgado, H. E. and Delisle, J.-B. and Di Matteo, P. and Diakite, S. and Diener, C. and Distefano, E. and Dolding, C. and Eappachen, D. and Edvardsson, B. and Enke, H. and Esquej, P. and Fabre, C. and Fabrizio, M. and Faigler, S. and Fedorets, G. and Fernique, P. and Fienga, A. and Figueras, F. and Fouron, C. and Fragkoudi, F. and Fraile, E. and Franke, F. and Gai, M. and Garabato, D. and Garcia-Gutierrez, A. and Garc{\'{i}}a-Torres, M. and Garofalo, A. and Gavras, P. and Gerlach, E. and Geyer, R. and Giacobbe, P. and Gilmore, G. and Girona, S. and Giuffrida, G. and Gomel, R. and Gomez, A. and Gonzalez-Santamaria, I. and Gonz{\'{a}}lez-Vidal, J. J. and Granvik, M. and Guti{\'{e}}rrez-S{\'{a}}nchez, R. and Guy, L. P. and Hauser, M. and Haywood, M. and Helmi, A. and Hidalgo, S. L. and Hilger, T. and H{\l}adczuk, N. and Hobbs, D. and Holland, G. and Huckle, H. E. and Jasniewicz, G. and Jonker, P. G. and Juaristi Campillo, J. and Julbe, F. and Karbevska, L. and Kervella, P. and Khanna, S. and Kochoska, A. and Kontizas, M. and Kordopatis, G. and Korn, A. J. and Kostrzewa-Rutkowska, Z. and Kruszy{\'{n}}ska, K. and Lambert, S. and Lanza, A. F. and Lasne, Y. and Le Campion, J.-F. and Le Fustec, Y. and Lebreton, Y. and Lebzelter, T. and Leccia, S. and Leclerc, N. and Lecoeur-Taibi, I. and Liao, S. and Licata, E. and Lindstr{\o}m, E. P. and Lister, T. A. and Livanou, E. and Lobel, A. and Madrero Pardo, P. and Managau, S. and Mann, R. G. and Marchant, J. M. and Marconi, M. and Marcos Santos, M. M. S. and Marinoni, S. and Marocco, F. and Marshall, D. J. and Martin Polo, L. and Mart{\'{i}}n-Fleitas, J. M. and Masip, A. and Massari, D. and Mastrobuono-Battisti, A. and Mazeh, T. and McMillan, P. J. and Messina, S. and Michalik, D. and Millar, N. R. and Mints, A. and Molina, D. and Molinaro, R. and Moln{\'{a}}r, L. and Montegriffo, P. and Mor, R. and Morbidelli, R. and Morel, T. and Morris, D. and Mulone, A. F. and Munoz, D. and Muraveva, T. and Murphy, C. P. and Musella, I. and Noval, L. and Ord{\'{e}}novic, C. and Orr{\`{u}}, G. and Osinde, J. and Pagani, C. and Pagano, I. and Palaversa, L. and Palicio, P. A. and Panahi, A. and Pawlak, M. and Pe{\~{n}}alosa Esteller, X. and Penttil{\"{a}}, A. and Piersimoni, A. M. and Pineau, F.-X. and Plachy, E. and Plum, G. and Poggio, E. and Poretti, E. and Poujoulet, E. and Pr{\v{s}}a, A. and Pulone, L. and Racero, E. and Ragaini, S. and Rainer, M. and Raiteri, C. M. and Rambaux, N. and Ramos, P. and Ramos-Lerate, M. and Re Fiorentin, P. and Regibo, S. and Reyl{\'{e}}, C. and Ripepi, V. and Riva, A. and Rixon, G. and Robichon, N. and Robin, C. and Roelens, M. and Rohrbasser, L. and Romero-G{\'{o}}mez, M. and Rowell, N. and Royer, F. and Rybicki, K. A. and Sadowski, G. and Sagrist{\`{a}} Sell{\'{e}}s, A. and Sahlmann, J. and Salgado, J. and Salguero, E. and Samaras, N. and Sanchez Gimenez, V. and Sanna, N. and Santove{\~{n}}a, R. and Sarasso, M. and Schultheis, M. and Sciacca, E. and Segol, M. and Segovia, J. C. and S{\'{e}}gransan, D. and Semeux, D. and Shahaf, S. and Siddiqui, H. I. and Siebert, A. and Siltala, L. and Slezak, E. and Smart, R. L. and Solano, E. and Solitro, F. and Souami, D. and Souchay, J. and Spagna, A. and Spoto, F. and Steele, I. A. and Steidelm{\"{u}}ller, H. and Stephenson, C. A. and S{\"{u}}veges, M. and Szabados, L. and Szegedi-Elek, E. and Taris, F. and Tauran, G. and Taylor, M. B. and Teixeira, R. and Thuillot, W. and Tonello, N. and Torra, F. and Torra, J. and Turon, C. and Unger, N. and Vaillant, M. and van Dillen, E. and Vanel, O. and Vecchiato, A. and Viala, Y. and Vicente, D. and Voutsinas, S. and Weiler, M. and Wevers, T. and Wyrzykowski, Ł. and Yoldas, A. and Yvard, P. and Zhao, H. and Zorec, J. and Zucker, S. and Zurbach, C. and Zwitter, T.},
    month = {5},
    pages = {A1},
    volume = {649},
    publisher = {EDP Sciences},
    url = {https://www.aanda.org/10.1051/0004-6361/202039657},
    doi = {10.1051/0004-6361/202039657},
    issn = {0004-6361}
}

@article{Faramaz2021A7,
    title = {{A Detailed Characterization of HR 8799's Debris Disk with ALMA in Band 7}},
    year = {2021},
    journal = {AJ},
    author = {Faramaz, Virginie and Marino, Sebastian and Booth, Mark and Matr{\`{a}}, Luca and Mamajek, Eric E. and Bryden, Geoffrey and Stapelfeldt, Karl R. and Casassus, Simon and Cuadra, Jorge and Hales, Antonio S. and Zurlo, Alice},
    number = {6},
    month = {6},
    pages = {271},
    volume = {161},
    url = {https://iopscience.iop.org/article/10.3847/1538-3881/abf4e0},
    doi = {10.3847/1538-3881/abf4e0},
    issn = {0004-6256}
}

@article{Nakatani2023AStars,
    title = {{A Primordial Origin for the Gas-rich Debris Disks around Intermediate-mass Stars}},
    year = {2023},
    journal = {ApJL},
    author = {Nakatani, Riouhei and Turner, Neal J. and Hasegawa, Yasuhiro and Cataldi, Gianni and Aikawa, Yuri and Marino, Sebastián and Kobayashi, Hiroshi},
    number = {2},
    month = {12},
    pages = {L28},
    volume = {959},
    publisher = {American Astronomical Society},
    doi = {10.3847/2041-8213/ad0ed8},
    issn = {2041-8205}
}

@article{Marshall2021AWavelengths,
    title = {{A search for trends in spatially resolved debris discs at far-infrared wavelengths}},
    year = {2021},
    journal = {MNRAS},
    author = {Marshall, J. P. and Wang, L. and Kennedy, G. M. and Zeegers, S. T. and Scicluna, P.},
    number = {4},
    month = {3},
    pages = {6168--6180},
    volume = {501},
    publisher = {Oxford University Press},
    doi = {10.1093/mnras/staa3917},
    issn = {13652966}
}

@article{Chen2012AScorpius-Centaurus,
    title = {{A Spitzer MIPS study of 2.5-2.0 M ⊙ stars in Scorpius-Centaurus}},
    year = {2012},
    journal = {ApJ},
    author = {Chen, Christine H. and Pecaut, Mark and Mamajek, Eric E. and Su, Kate Y.L. and Bitner, Martin},
    number = {2},
    month = {9},
    volume = {756},
    publisher = {Institute of Physics Publishing},
    doi = {10.1088/0004-637X/756/2/133},
    issn = {15384357}
}

@article{Yelverton2019ASystems,
    title = {{A statistically significant lack of debris discs in medium separation binary systems}},
    year = {2019},
    journal = {MNRAS},
    author = {Yelverton, Ben and Kennedy, Grant M. and Su, Kate Y.L. and Wyatt, Mark C.},
    number = {3},
    month = {9},
    pages = {3588--3606},
    volume = {488},
    publisher = {Oxford University Press},
    doi = {10.1093/mnras/stz1927},
    issn = {13652966},
    arxivId = {1907.04800}
}

@article{Pascucci2016ARELATION,
    title = {{A STEEPER THAN LINEAR DISK MASS–STELLAR MASS SCALING RELATION}},
    year = {2016},
    journal = {ApJ},
    author = {Pascucci, I and Testi, L and Herczeg, G J and Long, F and Manara, C F and Hendler, N and Mulders, G D and Krijt, S and Ciesla, F and Henning, Th and Mohanty, S and Drabek-Maunder, E and Apai, D and Sz{\H{u}}cs, L and Sacco, G and Olofsson, J},
    number = {2},
    pages = {125},
    volume = {831},
    url = {https://iopscience.iop.org/article/10.3847/0004-637X/831/2/125/pdf},
    doi = {10.3847/0004-637x/831/2/125},
    issn = {0004-637X},
    arxivId = {1608.03621}
}

@article{Pawellek2021AGroup,
    title = {{A ∼75 per cent occurrence rate of debris discs around F stars in the {$\beta$} Pic moving group}},
    year = {2021},
    journal = {MNRAS},
    author = {Pawellek, Nicole and Wyatt, Mark and Matra, Luca and Kennedy, Grant and Yelverton6, Ben},
    number = {4},
    month = {4},
    pages = {5390--5416},
    volume = {502},
    publisher = {Oxford University Press},
    doi = {10.1093/mnras/stab269},
    issn = {13652966}
}

@techreport{Cortes2025ALMAHandbook,
    title = {{ALMA Cycle 12 Technical Handbook}},
    year = {2025},
    author = {Cortes, P. C. and Carpenter, J. and Kameno, S. and Loomis, R. and Vila-Vilaro, B. and Immer, K. and Vlahakis, C. and Law, J. and Stoehr, F. and Saini, K. and Hales, A.},
    url = {https://doi.org/10.5281/zenodo.14933753},
    isbn = {9783923524662},
    doi = {10.5281/zenodo.14933753}
}

@article{TheAstropyCollaboration2013,
    title = {{Astropy: A Community Python Package for Astronomy}},
    year = {2013},
    journal = {A{\&}A},
    author = {{The Astropy Collaboration} and Robitaille, Thomas P. and Tollerud, Erik J. and Greenfield, Perry and Droettboom, Michael and Bray, Erik and Aldcroft, Tom and Davis, Matt and Ginsburg, Adam and Price-Whelan, Adrian M. and Kerzendorf, Wolfgang E. and Conley, Alexander and Crighton, Neil and Barbary, Kyle and Muna, Demitri and Ferguson, Henry and Grollier, Frédéric and Parikh, Madhura M. and Nair, Prasanth H. and G{\"{u}}nther, Hans M. and Deil, Christoph and Woillez, Julien and Conseil, Simon and Kramer, Roban and Turner, James E. H. and Singer, Leo and Fox, Ryan and Weaver, Benjamin A. and Zabalza, Victor and Edwards, Zachary I. and Bostroem, K. Azalee and Burke, D. J. and Casey, Andrew R. and Crawford, Steven M. and Dencheva, Nadia and Ely, Justin and Jenness, Tim and Labrie, Kathleen and Lim, Pey Lian and Pierfederici, Francesco and Pontzen, Andrew and Ptak, Andy and Refsdal, Brian and Servillat, Mathieu and Streicher, Ole},
    month = {7},
    pages = {A33},
    volume = {558},
    url = {http://arxiv.org/abs/1307.6212},
    doi = {10.1051/0004-6361/201322068},
    arxivId = {1307.6212}
}

@article{Dent2005,
    title = {{CO emission from discs around isolated HAeBe and Vega-excess stars}},
    year = {2005},
    journal = {MNRAS},
    author = {Dent, W. R. F. and Greaves, J. S. and Coulson, I. M.},
    number = {2},
    month = {5},
    pages = {663--676},
    volume = {359},
    publisher = {Oxford University Press},
    url = {https://academic.oup.com/mnras/article-lookup/doi/10.1111/j.1365-2966.2005.08938.x},
    doi = {10.1111/j.1365-2966.2005.08938.x}
}

@article{Marino2017ALMAPlanets,
    title = {{ALMA observations of the {$\eta$} Corvi debris disc: inward scattering of CO-rich exocomets by a chain of 3-30 M ⊕ planets?}},
    year = {2017},
    journal = {MNRAS},
    author = {Marino, S and Wyatt, M C and Pani´c, O and Matr{\`{a}}, L and Kennedy, G M and Bonsor, A and Kral, Q and Dent, W R F and Duchene, G and Wilner, D and Lisse, C M and Lestrade, J.-F and Matthews, B},
    pages = {2595--2615},
    volume = {465},
    url = {https://almascience.eso.org/news/notification-of-problem-},
    doi = {10.1093/mnras/stw2867}
}

@article{Mayama2018ALMADisk,
    title = {{ALMA Reveals a Misaligned Inner Gas Disk inside the Large Cavity of a Transitional Disk}},
    year = {2018},
    journal = {ApJ},
    author = {Mayama, Satoshi and Akiyama, Eiji and Pani{\'{c}}, Olja and Miley, James and Tsukagoshi, Takashi and Muto, Takayuki and Dong, Ruobing and de Leon, Jerome and Mizuki, Toshiyuki and Oh, Daehyeon and Hashimoto, Jun and Sai, Jinshi and Currie, Thayne and Takami, Michihiro and Grady, Carol A. and Hayashi, Masahiko and Tamura, Motohide and Inutsuka, Shu-ichiro},
    number = {1},
    month = {11},
    pages = {L3},
    volume = {868},
    url = {http://stacks.iop.org/2041-8205/868/i=1/a=L3?key=crossref.4a954399bde606e8a6195c2fd774e9a5 http://arxiv.org/abs/1810.06941%0Ahttp://dx.doi.org/10.3847/2041-8213/aae88b},
    doi = {10.3847/2041-8213/aae88b},
    arxivId = {1810.06941}
}

@article{Hogbom1974ApertureBaselines,
    title = {{Aperture Synthesis with a Non-Regular Distribution of Interferometer Baselines}},
    year = {1974},
    journal = {A\&A Supplement Series},
    author = {H{\"{o}}gbom, J. A.},
    month = {6},
    pages = {417},
    volume = {15},
    url = {https://ui.adsabs.harvard.edu/abs/1974A%26AS...15..417H/abstract},
    issn = {0365-0138,0004-6361}
}

@article{Harris2020ArrayNumPy,
    title = {{Array programming with NumPy}},
    year = {2020},
    journal = {Nature},
    author = {Harris, Charles R. and Millman, K. Jarrod and van der Walt, Stéfan J. and Gommers, Ralf and Virtanen, Pauli and Cournapeau, David and Wieser, Eric and Taylor, Julian and Berg, Sebastian and Smith, Nathaniel J. and Kern, Robert and Picus, Matti and Hoyer, Stephan and van Kerkwijk, Marten H. and Brett, Matthew and Haldane, Allan and del R{\'{i}}o, Jaime Fernández and Wiebe, Mark and Peterson, Pearu and G{\'{e}}rard-Marchant, Pierre and Sheppard, Kevin and Reddy, Tyler and Weckesser, Warren and Abbasi, Hameer and Gohlke, Christoph and Oliphant, Travis E.},
    number = {7825},
    month = {9},
    pages = {357--362},
    volume = {585},
    publisher = {Nature Research},
    url = {https://www.nature.com/articles/s41586-020-2649-2},
    doi = {10.1038/s41586-020-2649-2},
    issn = {0028-0836}
}

@article{Kepley2020Auto-multithresh:Algorithm,
    title = {{auto-multithresh: A General Purpose Automasking Algorithm}},
    year = {2020},
    journal = {PASP},
    author = {Kepley, Amanda A. and Tsutsumi, Takahiro and Brogan, Crystal L. and Indebetouw, Remy and Yoon, Ilsang and Mason, Brian and Meyer, Jennifer Donovan},
    number = {1008},
    month = {2},
    pages = {024505},
    volume = {132},
    publisher = {Institute of Physics Publishing},
    url = {https://iopscience.iop.org/article/10.1088/1538-3873/ab5e14},
    doi = {10.1088/1538-3873/ab5e14},
    issn = {0004-6280}
}

@article{Spilker2022BirdsScales,
    title = {{Bird’s eye view of molecular clouds in the Milky Way - II. Cloud kinematics from subparsec to kiloparsec scales}},
    year = {2022},
    journal = {A{\&}A},
    author = {Spilker, Andri and Kainulainen, Jouni and Orkisz, Jan},
    month = {11},
    pages = {A110},
    volume = {667},
    publisher = {EDP Sciences},
    url = {https://www.aanda.org/articles/aa/full_html/2022/11/aa44392-22/aa44392-22.html https://www.aanda.org/articles/aa/abs/2022/11/aa44392-22/aa44392-22.html},
    doi = {10.1051/0004-6361/202244392},
    issn = {0004-6361}
}

@article{Matthews2018,
    title = {{Constraining the presence of giant planets in two-belt debris disc systems with VLT/SPHERE direct imaging and dynamical arguments}},
    year = {2018},
    journal = {MNRAS},
    author = {Matthews, Elisabeth and Hinkley, Sasha and Vigan, Arthur and Kennedy, Grant and Sutlieff, Ben and Wickenden, Dawn and Treves, Sam and David, Trevor and Meshkat, Tiffany and Mawet, Dimitri and Morales, Farisa and Shannon, Andrew and Stapelfeldt, Karl},
    number = {2},
    month = {10},
    pages = {2757--2783},
    volume = {480},
    url = {https://academic.oup.com/mnras/article/480/2/2757/5049325},
    doi = {10.1093/mnras/sty1778},
    issn = {0035-8711}
}

@article{Lieman-Sifry2016,
    title = {{DEBRIS DISKS IN THE SCORPIUS–CENTAURUS OB ASSOCIATION RESOLVED BY ALMA}},
    year = {2016},
    journal = {ApJ},
    author = {Lieman-Sifry, Jesse and Hughes, A. Meredith and Carpenter, John M. and Gorti, Uma and Hales, Antonio and Flaherty, Kevin M.},
    number = {1},
    month = {8},
    pages = {25},
    volume = {828},
    publisher = {IOP Publishing},
    url = {http://stacks.iop.org/0004-637X/828/i=1/a=25?key=crossref.5c45a641dbd32b5932267e5ae1dc370d},
    doi = {10.3847/0004-637X/828/1/25},
    issn = {1538-4357}
}

@article{Marino2016,
    title = {{Exocometary gas in the HD 181327 debris ring}},
    year = {2016},
    journal = {MNRAS},
    author = {Marino, S. and Matr{\`{a}}, L. and Stark, C. and Wyatt, M. C. and Casassus, S. and Kennedy, G. and Rodriguez, D. and Zuckerman, B. and Perez, S. and Dent, W. R. F. and Kuchner, M. and Hughes, A. M. and Schneider, G. and Steele, A. and Roberge, A. and Donaldson, J. and Nesvold, E.},
    number = {3},
    month = {8},
    pages = {2933--2944},
    volume = {460},
    publisher = {Oxford University Press},
    url = {http://mnras.oxfordjournals.org/lookup/doi/10.1093/mnras/stw1216},
    doi = {10.1093/mnras/stw1216},
    issn = {0035-8711}
}

@article{Stapper2024ConstrainingIsotopologues,
    title = {{Constraining the gas mass of Herbig disks using CO isotopologues}},
    year = {2024},
    journal = {A\&A},
    author = {Stapper, L. M. and Hogerheijde, M. R. and van Dishoeck, E. F. and Lin, L. and Ahmadi, A. and Booth, A. S. and Grant, S. L. and Immer, K. and Leemker, M. and P{\'{e}}rez-S{\'{a}}nchez, A. F.},
    month = {2},
    volume = {682},
    publisher = {EDP Sciences},
    doi = {10.1051/0004-6361/202347271},
    issn = {14320746},
    arxivId = {2312.03835}
}

@article{Chabrier2000CoolingAtmospheres,
    title = {{Cooling Sequences and Color‐Magnitude Diagrams for Cool White Dwarfs with Hydrogen Atmospheres}},
    year = {2000},
    journal = {ApJ},
    author = {Chabrier, G. and Brassard, P. and Fontaine, G. and Saumon, D.},
    number = {1},
    month = {11},
    pages = {216--226},
    volume = {543},
    url = {https://iopscience.iop.org/article/10.1086/317092},
    doi = {10.1086/317092},
    issn = {0004-637X}
}

@article{Raymond2011DebrisFormation,
    title = {{Debris disks as signposts of terrestrial planet formation}},
    year = {2011},
    journal = {A\&A},
    author = {Raymond, S. N. and Armitage, P. J. and Moro-Mart{\'{i}}n, A. and Booth, M. and Wyatt, M. C. and Armstrong, J. C. and Mandell, A. M. and Selsis, F. and West, A. A.},
    volume = {530},
    doi = {10.1051/0004-6361/201116456},
    issn = {00046361}
}

@article{Matra2017DetectionComets,
    title = {{Detection of Exocometary CO within the 440 Myr Old Fomalhaut Belt: A Similar CO+CO 2 Ice Abundance in Exocomets and Solar System Comets}},
    year = {2017},
    journal = {ApJ},
    author = {Matr{\`{a}}, L. and MacGregor, M. A. and Kalas, P. and Wyatt, M. C. and Kennedy, G. M. and Wilner, D. J. and Duchene, G. and Hughes, A. M. and Pan, M. and Shannon, A. and Clampin, M. and Fitzgerald, M. P. and Graham, J. R. and Holland, W. S. and Pani{\'{c}}, O. and Su, K. Y. L.},
    number = {1},
    month = {6},
    pages = {9},
    volume = {842},
    url = {https://iopscience.iop.org/article/10.3847/1538-4357/aa71b4},
    doi = {10.3847/1538-4357/aa71b4},
    issn = {0004-637X},
    arxivId = {1705.05868}
}

@article{Marois2008Direct8799.,
    title = {{Direct imaging of multiple planets orbiting the star HR 8799.}},
    year = {2008},
    journal = {Science},
    author = {Marois, Christian and Macintosh, Bruce and Barman, Travis and Zuckerman, B and Song, Inseok and Patience, Jennifer and Lafreni{\`{e}}re, David and Doyon, René},
    number = {5906},
    month = {10},
    pages = {1348--1352},
    volume = {322},
    publisher = {American Association for the Advancement of Science},
    url = {http://www.ncbi.nlm.nih.gov/pubmed/19008415 https://www.sciencemag.org/lookup/doi/10.1126/science.1166585},
    doi = {10.1126/science.1166585},
    issn = {1095-9203},
    pmid = {19008415}
}

@article{Ribas2014DiskMyr,
    title = {{Disk evolution in the solar neighborhood: I. Disk frequencies from 1 to 100 Myr}},
    year = {2014},
    journal = {A\&A},
    author = {Ribas, Alvaro and Mer{\'{i}}n, Bruno and Bouy, Hervé and Maud, Luke T.},
    month = {1},
    volume = {561},
    doi = {10.1051/0004-6361/201322597},
    issn = {00046361},
    arxivId = {1312.0609}
}

@article{Kennedy2014DoBelts,
    title = {{Do two-temperature debris discs have multiple belts?}},
    year = {2014},
    journal = {MNRAS},
    author = {Kennedy, G M and Wyatt, M C},
    number = {4},
    month = {10},
    pages = {3164--3182},
    volume = {444},
    publisher = {Oxford University Press},
    url = {http://academic.oup.com/mnras/article/444/4/3164/1021189/Do-twotemperature-debris-discs-have-multiple-belts},
    doi = {10.1093/mnras/stu1665}
}

@article{Matra2020DustALMA,
    title = {{Dust Populations in the Iconic Vega Planetary System Resolved by ALMA}},
    year = {2020},
    journal = {ApJ},
    author = {Matr{\`{a}}, Luca and Dent, William R F and Wilner, David J and Marino, Sebastián and Wyatt, Mark C and Marshall, Jonathan P and Su, Kate Y L and Chavez, Miguel and Hales, Antonio and Hughes, A Meredith and Greaves, Jane S and Corder, Stuartt A},
    pages = {146},
    volume = {898},
    url = {https://doi.org/10.3847/1538-4357/aba0a4},
    doi = {10.3847/1538-4357/aba0a4}
}

@article{Pearce2014DynamicalDisc,
    title = {{Dynamical evolution of an eccentric planet and a less massive debris disc}},
    year = {2014},
    author = {Pearce, Tim D. and Wyatt, Mark C.},
    month = {6},
    url = {http://arxiv.org/abs/1406.7294 http://dx.doi.org/10.1093/mnras/stu1302},
    doi = {10.1093/mnras/stu1302},
    arxivId = {1406.7294}
}

@article{Lynch2022EccentricDiscs,
    title = {{Eccentric debris disc morphologies - I. Exploring the origin of apocentre and pericentre glows in face-on debris discs}},
    year = {2022},
    journal = {MNRAS},
    author = {Lynch, Elliot M. and Lovell, Joshua B.},
    number = {2},
    month = {2},
    pages = {2538--2551},
    volume = {510},
    publisher = {Oxford University Press},
    doi = {10.1093/mnras/stab3566},
    issn = {13652966}
}

@article{Lovell2023EccentricDiscs,
    title = {{Eccentric debris disc morphologies - II. Surface brightness variations from overlapping orbits in narrow eccentric discs}},
    year = {2023},
    journal = {MNRASL},
    author = {Lovell, Joshua B. and Lynch, Elliot M.},
    number = {1},
    month = {10},
    pages = {L36-L42},
    volume = {525},
    publisher = {Oxford University Press},
    doi = {10.1093/mnrasl/slad083},
    issn = {17453933}
}

@article{Foreman-Mackey2013EmceeHammer,
    title = {{emcee : The MCMC Hammer}},
    year = {2013},
    journal = {PASP},
    author = {Foreman-Mackey, Daniel and Hogg, David W and Lang, Dustin and Goodman, Jonathan},
    number = {925},
    pages = {306--312},
    volume = {125},
    url = {http://dan.iel.fm/emceeundertheMITLicense.},
    doi = {10.1086/670067},
    issn = {00046280},
    arxivId = {1202.3665}
}

@article{Bailer-Jones2021Estimating3,
    title = {{Estimating Distances from Parallaxes. V. Geometric and Photogeometric Distances to 1.47 Billion Stars in Gaia Early Data Release 3}},
    year = {2021},
    journal = {AJ},
    author = {Bailer-Jones, C. A. L. and Rybizki, J. and Fouesneau, M. and Demleitner, M. and Andrae, R.},
    number = {3},
    pages = {147},
    volume = {161},
    publisher = {IOP Publishing},
    url = {http://dx.doi.org/10.3847/1538-3881/abd806},
    doi = {10.3847/1538-3881/abd806},
    issn = {0004-6256},
    arxivId = {2012.05220}
}

@article{Lagrange2019EvidenceSystem,
    title = {{Evidence for an additional planet in the {$\beta$} Pictoris system}},
    year = {2019},
    journal = {Nature Ast},
    author = {Lagrange, A.-M. and Meunier, Nadège and Rubini, Pascal and Keppler, Miriam and Galland, Franck and Chapellier, Eric and Michel, Eric and Balona, Luis and Beust, Hervé and Guillot, Tristan and Grandjean, Antoine and Borgniet, Simon and M{\'{e}}karnia, Djamel and Wilson, Paul Anthony and Kiefer, Flavien and Bonnefoy, Mickael and Lillo-Box, Jorge and Pantoja, Blake and Jones, Matias and Iglesias, Daniela Paz and Rodet, Laetitia and Diaz, Matias and Zapata, Abner and Abe, Lyu and Schmider, François-Xavier},
    number = {12},
    pages = {1135--1142},
    volume = {3},
    url = {https://doi.org/10.1038/s41550-019-0857-1},
    doi = {10.1038/s41550-019-0857-1},
    issn = {2397-3366}
}

@article{Wyatt2008EvolutionDisks,
    title = {{Evolution of debris disks}},
    year = {2008},
    journal = {Annual Review of A\&A},
    author = {Wyatt, Mark C},
    pages = {339--383},
    volume = {46},
    url = {www.annualreviews.org},
    isbn = {9780824309466},
    doi = {10.1146/annurev.astro.45.051806.110525},
    issn = {00664146}
}

@article{Baraffe2003Evolutionary209458,
    title = {{Evolutionary models for cool brown dwarfs and extrasolar giant planets. The case of HD 209458}},
    year = {2003},
    journal = {A{\&}A},
    author = {Baraffe, I. and Chabrier, G. and Barman, T. S. and Allard, F. and Hauschildt, P. H.},
    number = {2},
    pages = {701--712},
    volume = {402},
    publisher = {EDP Sciences},
    doi = {10.1051/0004-6361:20030252},
    issn = {00046361},
    arxivId = {astro-ph/0302293}
}

@article{Yang2024FirstStar,
    title = {{First ALMA observations of the HD 105211 debris disc: A warm dust component close to a gigayear-old star}},
    year = {2024},
    journal = {A{\&}A},
    author = {Yang, Qiancheng and Liu, Qiong and Kennedy, Grant M. and Wyatt, Mark C. and Dodson-Robinson, Sarah and Akeson, Rachel and Liao, Nenghui},
    month = {6},
    pages = {A206},
    volume = {686},
    publisher = {EDP Sciences},
    url = {https://ui.adsabs.harvard.edu/abs/2024A%26A...686A.206Y/abstract},
    doi = {10.1051/0004-6361/202449280},
    issn = {0004-6361}
}

@article{Wyatt2015,
    title = {{Five steps in the evolution from protoplanetary to debris disk}},
    year = {2015},
    journal = {ApASS},
    author = {Wyatt, M. C. and Pani{\'{c}}, O. and Kennedy, G. M. and Matr{\`{a}}, L.},
    number = {2},
    month = {6},
    pages = {103},
    volume = {357},
    publisher = {Springer Netherlands},
    url = {http://link.springer.com/10.1007/s10509-015-2315-6},
    doi = {10.1007/s10509-015-2315-6},
    issn = {0004-640X}
}

@article{Cuello2020FlybysSignatures,
    title = {{Flybys in protoplanetary discs - II. Observational signatures}},
    year = {2020},
    journal = {MNRAS},
    author = {Cuello, Nicolás and Louvet, Fabien and Mentiplay, Daniel and Pinte, Christophe and Price, Daniel J and Winter, Andrew J and Nealon, Rebecca and M{\'{e}}nard, François and Lodato, Giuseppe and Dipierro, Giovanni and Christiaens, Valentin and Montesinos, Matías and Cuadra, Jorge and Laibe, Guillaume and Cieza, Lucas and Dong, Ruobing and Alexander, Richard},
    number = {1},
    month = {1},
    pages = {504--514},
    volume = {491},
    url = {https://arxiv.org/pdf/1910.06822.pdf https://academic.oup.com/mnras/article/491/1/504/5601768},
    doi = {10.1093/mnras/stz2938},
    issn = {13652966},
    arxivId = {1910.06822}
}

@article{Kral2020FormationAccretion,
    title = {{Formation of secondary atmospheres on terrestrial planets by late disk accretion}},
    year = {2020},
    journal = {Nature Ast},
    author = {Kral, Quentin and Davoult, Jeanne and Charnay, Benjamin},
    number = {8},
    month = {8},
    pages = {769--775},
    volume = {4},
    url = {http://arxiv.org/abs/2004.02496 http://dx.doi.org/10.1038/s41550-020-1050-2 http://www.nature.com/articles/s41550-020-1050-2},
    doi = {10.1038/s41550-020-1050-2},
    issn = {2397-3366},
    arxivId = {2004.02496}
}

@article{Friebe2022GapDisc,
    title = {{Gap carving by a migrating planet embedded in a massive debris disc}},
    year = {2022},
    journal = {MNRAS},
    author = {Friebe, Marc F and Pearce, Tim D and L{\"{o}}hne, Torsten},
    number = {3},
    month = {4},
    pages = {4441--4454},
    volume = {512},
    url = {https://academic.oup.com/mnras/article/512/3/4441/6547020},
    doi = {10.1093/mnras/stac664},
    issn = {0035-8711}
}

@article{Trapman2019GasEvolution,
    title = {{Gas versus dust sizes of protoplanetary discs: effects of dust evolution}},
    year = {2019},
    journal = {A{\&}A},
    author = {Trapman, L and Facchini, S and Hogerheijde, M R and van Dishoeck, E. F. and Bruderer, S},
    pages = {A79},
    volume = {629},
    url = {https://arxiv.org/pdf/1903.06190.pdf},
    doi = {10.1051/0004-6361/201834723},
    issn = {0004-6361},
    arxivId = {1903.06190}
}

@article{Johnson2010GiantPlane,
    title = {{Giant Planet Occurrence in the Stellar Mass-Metallicity Plane}},
    year = {2010},
    journal = {PASP},
    author = {Johnson, John Asher and Aller, Kimberly M and Howard, Andrew W and Crepp, Justin R},
    number = {894},
    month = {10},
    pages = {905--915},
    volume = {122},
    url = {http://iopscience.iop.org/article/10.1086/655775},
    doi = {10.1086/655775},
    issn = {0004-6280}
}

@article{Acke2012HerschelActivity,
    title = {{Herschel images of Fomalhaut - An extrasolar Kuiper belt at the height of its dynamical activity}},
    year = {2012},
    journal = {A{\&}A},
    author = {Acke, B. and Min, M. and Dominik, C. and Vandenbussche, B. and Sibthorpe, B. and Waelkens, C. and Olofsson, G. and Degroote, P. and Smolders, K. and Pantin, E. and Barlow, M. J. and Blommaert, J. A.D.L. and Brandeker, A. and De Meester, W. and Dent, W. R.F. and Exter, K. and Di Francesco, J. and Fridlund, M. and Gear, W. K. and Glauser, A. M. and Greaves, J. S. and Harvey, P. M. and Henning, Th and Hogerheijde, M. R. and Holland, W. S. and Huygen, R. and Ivison, R. J. and Jean, C. and Liseau, R. and Naylor, D. A. and Pilbratt, G. L. and Polehampton, E. T. and Regibo, S. and Royer, P. and Sicilia-Aguilar, A. and Swinyard, B. M.},
    month = {4},
    pages = {A125},
    volume = {540},
    publisher = {EDP Sciences},
    url = {https://www.aanda.org/articles/aa/full_html/2012/04/aa18581-11/aa18581-11.html https://www.aanda.org/articles/aa/abs/2012/04/aa18581-11/aa18581-11.html},
    doi = {10.1051/0004-6361/201118581},
    issn = {0004-6361}
}

@article{Marois2010Images8799,
    title = {{Images of a fourth planet orbiting HR 8799}},
    year = {2010},
    journal = {Nature},
    author = {Marois, Christian and Zuckerman, B and Konopacky, Quinn M and Macintosh, Bruce and Barman, Travis},
    number = {7327},
    pages = {1080--1083},
    volume = {468},
    url = {https://arxiv.org/pdf/1011.4918.pdf},
    doi = {10.1038/nature09684},
    issn = {00280836},
    arxivId = {1011.4918}
}

@article{Kral2019ImagingDiscs,
    title = {{Imaging [CI] around HD 131835: reinterpreting young debris discs with protoplanetary disc levels of CO gas as shielded secondary discs}},
    year = {2019},
    journal = {MNRAS},
    author = {Kral, Quentin and Marino, Sebastian and Wyatt, Mark C and Kama, Mihkel and Matr{\'{a}}, Luca},
    number = {3},
    month = {11},
    pages = {3670--3691},
    volume = {489},
    publisher = {Narnia},
    url = {https://academic.oup.com/mnras/advance-article/doi/10.1093/mnras/sty2923/5185097},
    doi = {10.1093/mnras/sty2923},
    issn = {0035-8711}
}

@article{Gelman1992InferenceSequences,
    title = {{Inference from Iterative Simulation Using Multiple Sequences}},
    year = {1992},
    journal = {https://doi.org/10.1214/ss/1177011136},
    author = {Gelman, Andrew and Rubin, Donald B.},
    number = {4},
    month = {11},
    pages = {457--472},
    volume = {7},
    publisher = {Institute of Mathematical Statistics},
    url = {https://projecteuclid.org/journals/statistical-science/volume-7/issue-4/Inference-from-Iterative-Simulation-Using-Multiple-Sequences/10.1214/ss/1177011136.full https://projecteuclid.org/journals/statistical-science/volume-7/issue-4/Inference-from-Iterative-Simulation-Using-Multiple-Sequences/10.1214/ss/1177011136.short},
    doi = {10.1214/SS/1177011136},
    issn = {0883-4237}
}

@article{Marino2020InsightsALMA,
    title = {{Insights into the planetary dynamics of HD 206893 with ALMA}},
    year = {2020},
    journal = {MNRAS},
    author = {Marino, S and Zurlo, A and Faramaz, V and Milli, J and Henning, Th and Kennedy, G M and Matr{\`{a}}, L and P{\'{e}}rez, S and Delorme, P and Cieza, L A and Hughes, A M},
    number = {1},
    month = {10},
    pages = {1319--1334},
    volume = {498},
    url = {https://academic.oup.com/mnras/article/498/1/1319/5899754},
    doi = {10.1093/mnras/staa2386},
    issn = {0035-8711}
}

@article{Launhardt2020ISPY-NACOStars,
    title = {{ISPY-NACO Imaging Survey for Planets around Young stars}},
    year = {2020},
    journal = {A{\&}A},
    author = {Launhardt, R. and Henning, Th. and Quirrenbach, A. and S{\'{e}}gransan, D. and Avenhaus, H. and van Boekel, R. and Brems, S. S. and Cheetham, A. C. and Cugno, G. and Girard, J. and Godoy, N. and Kennedy, G. M. and Maire, A.-L. and Metchev, S. and M{\"{u}}ller, A. and Musso Barcucci, A. and Olofsson, J. and Pepe, F. and Quanz, S. P. and Queloz, D. and Reffert, S. and Rickman, E. L. and Ruh, H. L. and Samland, M.},
    month = {3},
    pages = {A162},
    volume = {635},
    publisher = {EDP Sciences},
    url = {https://www.aanda.org/10.1051/0004-6361/201937000},
    doi = {10.1051/0004-6361/201937000},
    issn = {0004-6361}
}

@article{Rebollido2024JWST-TSTMIRI,
    title = {{JWST-TST High Contrast: Asymmetries, dust populations and hints of a collision in the {\$}{$\beta$}{\$} Pictoris disk with NIRCam and MIRI}},
    year = {2024},
    author = {Rebollido, Isabel and Stark, Christopher C. and Kammerer, Jens and Perrin, Marshall D. and Lawson, Kellen and Pueyo, Laurent and Chen, Christine and Hines, Dean and Girard, Julien H. and Worthen, Kadin and Ingerbretsen, Carl and Betti, Sarah and Clampin, Mark and Golimowski, David and Hoch, Kielan and Lewis, Nikole K. and Lu, Cicero X. and van der Marel, Roeland P. and Rickman, Emily and Seager, Sara and Soummer, Remi and Valenti, Jeff A. and Ward-Duong, Kimberly and Mountain, C. Matt},
    month = {1},
    url = {http://arxiv.org/abs/2401.05271},
    arxivId = {2401.05271}
}

@article{Matra2019KuiperALMA,
    title = {{Kuiper Belt–like Hot and Cold Populations of Planetesimal Inclinations in the {$\beta$} Pictoris Belt Revealed by ALMA}},
    year = {2019},
    journal = {AJ},
    author = {Matr{\`{a}}, L and Wyatt, M C and Wilner, D J and Dent, W R F and Marino, S and Kennedy, G M and Milli, J},
    number = {4},
    pages = {135},
    volume = {157},
    url = {https://arxiv.org/pdf/1902.04081.pdf},
    doi = {10.3847/1538-3881/ab06c0},
    issn = {1538-3881},
    arxivId = {1902.04081}
}

@article{Moor2017,
    title = {{Molecular Gas in Debris Disks around Young A-type Stars}},
    year = {2017},
    journal = {ApJ},
    author = {Mo{\'{o}}r, Attila and Cur{\'{e}}, Michel and K{\'{o}}sp{\'{a}}l, Agnes and Abrah{\'{a}}m, Péter and Csengeri, Timea and Eiroa, Carlos and Gunawan, Diah and Henning, Thomas and Hughes, A Meredith and Juh{\'{a}}sz, Attila and Pawellek, Nicole and Wyatt, Mark},
    number = {2},
    month = {11},
    pages = {123},
    volume = {849},
    url = {https://doi.org/10.3847/1538-4357/aa8e4e c},
    doi = {10.3847/1538-4357/aa8e4e},
    issn = {1538-4357}
}

@article{Tazzari2016,
    title = {{Multiwavelength analysis for interferometric (sub-)mm observations of protoplanetary disks}},
    year = {2016},
    journal = {A{\&}A},
    author = {Tazzari, M. and Testi, L. and Ercolano, B. and Natta, A. and Isella, A. and Chandler, C. J. and P{\'{e}}rez, L. M. and Andrews, S. and Wilner, D. J. and Ricci, L. and Henning, T. and Linz, H. and Kwon, W. and Corder, S. A. and Dullemond, C. P. and Carpenter, J. M. and Sargent, A. I. and Mundy, L. and Storm, S. and Calvet, N. and Greaves, J. A. and Lazio, J. and Deller, A. T.},
    month = {4},
    pages = {A53},
    volume = {588},
    publisher = {EDP Sciences},
    url = {http://www.aanda.org/10.1051/0004-6361/201527423},
    doi = {10.1051/0004-6361/201527423},
    issn = {0004-6361}
}

@article{MacGregor2013MillimeterDisk,
    title = {{Millimeter emission structure in the first alma image of the AU Mic debris disk}},
    year = {2013},
    journal = {APJL},
    author = {MacGregor, Meredith A. and Wilner, David J. and Rosenfeld, Katherine A. and Andrews, Sean M. and Matthews, Brenda and Hughes, A. Meredith and Booth, Mark and Chiang, Eugene and Graham, James R. and Kalas, Paul and Kennedy, Grant and Sibthorpe, Bruce},
    number = {2},
    month = {1},
    volume = {762},
    doi = {10.1088/2041-8205/762/2/L21},
    issn = {20418205},
    arxivId = {1211.5148}
}

@article{Dent2014MolecularDisk,
    title = {{Molecular gas clumps from the destruction of icy bodies in the {$\beta$} pictoris debris disk}},
    year = {2014},
    journal = {Science},
    author = {Dent, W. R.F. and Wyatt, M C and Roberge, A and Augereau, J. C. and Casassus, S and Corder, S and Greaves, J S and De Gregorio-Monsalvo, I. and Hales, A and Jackson, A P and Meredith Hughes, A. and Lagrange, A. M. and Matthews, B and Wilner, D},
    number = {6178},
    month = {3},
    pages = {1490--1492},
    volume = {343},
    publisher = {American Association for the Advancement of Science},
    url = {http://www.ncbi.nlm.nih.gov/pubmed/24603151},
    doi = {10.1126/science.1248726},
    issn = {10959203},
    pmid = {24603151}
}

@article{Sun2020MolecularPopulation,
    title = {{Molecular Gas Properties on Cloud Scales across the Local Star-forming Galaxy Population}},
    year = {2020},
    journal = {ApJL},
    author = {Sun, Jiayi and Leroy, Adam K. and Schinnerer, Eva and Hughes, Annie and Rosolowsky, Erik and Querejeta, Miguel and Schruba, Andreas and Liu, Daizhong and Saito, Toshiki and Herrera, Cinthya N. and Faesi, Christopher and Usero, Antonio and Pety, Jérôme and Kruijssen, J. M. Diederik and Ostriker, Eve C. and Bigiel, Frank and Blanc, Guillermo A. and Bolatto, Alberto D. and Boquien, Médéric and Chevance, Mélanie and Dale, Daniel A. and Deger, Sinan and Emsellem, Eric and Glover, Simon C. O. and Grasha, Kathryn and Groves, Brent and Henshaw, Jonathan and Jimenez-Donaire, Maria J. and Kim, Jenny J. and Klessen, Ralf S. and Kreckel, Kathryn and Lee, Janice C. and Meidt, Sharon and Sandstrom, Karin and Sardone, Amy E. and Utomo, Dyas and Williams, Thomas G.},
    number = {1},
    month = {9},
    pages = {L8},
    volume = {901},
    publisher = {IOP Publishing},
    url = {https://iopscience.iop.org/article/10.3847/2041-8213/abb3be https://iopscience.iop.org/article/10.3847/2041-8213/abb3be/meta},
    doi = {10.3847/2041-8213/ABB3BE},
    issn = {2041-8205},
    arxivId = {2009.01842}
}

@article{Tazzari2017Mtazzari/uvplot,
    title = {{mtazzari/uvplot}},
    year = {2017},
    journal = {Zenodo},
    author = {Tazzari, Marco},
    month = {10},
    pages = {10.5281/zenodo.1003113},
    volume = {v0.1.1},
    publisher = {Zenodo},
    url = {https://doi.org/10.5281/zenodo.1003113},
    doi = {10.5281/zenodo.1003113}
}

@article{Ida2000OrbitalObjects,
    title = {{Orbital Migration of Neptune and Orbital Distribution of Trans-Neptunian Objects}},
    year = {2000},
    journal = {ApJ},
    author = {Ida, Shigeru and Bryden, Geoffrey and Lin, D N C and Tanaka, Hidekazu},
    pages = {428--445},
    volume = {534},
    url = {www.harvard.edu/8}
}

@article{Reffert2014,
    title = {{Precise Radial Velocities of Giant Stars VII. Occurrence Rate of Giant Extrasolar Planets as a Function of Mass and Metallicity}},
    year = {2014},
    journal = {A{\&}A},
    author = {Reffert, Sabine and Bergmann, Christoph and Quirrenbach, Andreas and Trifonov, Trifon and K{\"{u}}nstler, Andreas},
    pages = {116},
    volume = {574},
    url = {http://www.exoplanets.org/ http://arxiv.org/abs/1412.4634%0Ahttp://dx.doi.org/10.1051/0004-6361/201322360},
    doi = {10.1051/0004-6361/201322360},
    arxivId = {1412.4634}
}

@article{Booth2013,
    title = {{Resolved debris discs around A stars in the Herschel DEBRIS survey}},
    year = {2013},
    journal = {MNRAS},
    author = {Booth, M. and Kennedy, G. and Sibthorpe, B. and Matthews, B. C. and Wyatt, M. C. and Duchene, G. and Kavelaars, J. J. and Rodriguez, D. and Greaves, J. S. and Koning, A. and Vican, L. and Rieke, G. H. and Su, K. Y. L. and Moro-Martin, A. and Kalas, P.},
    number = {2},
    month = {1},
    pages = {1263--1280},
    volume = {428},
    url = {https://academic.oup.com/mnras/article-lookup/doi/10.1093/mnras/sts117},
    doi = {10.1093/mnras/sts117},
    issn = {0035-8711}
}

@article{Nakatani2021PhotoevaporationRemnants,
    title = {{Photoevaporation of Grain-depleted Protoplanetary Disks around Intermediate-mass Stars: Investigating the Possibility of Gas-rich Debris Disks as Protoplanetary Remnants}},
    year = {2021},
    journal = {ApJ},
    author = {Nakatani, Riouhei and Kobayashi, Hiroshi and Kuiper, Rolf and Nomura, Hideko and Aikawa, Yuri},
    number = {2},
    pages = {90},
    volume = {915},
    doi = {10.3847/1538-4357/ac0137},
    issn = {0004-637X},
    arxivId = {2009.06438}
}

@article{Pearce2022PlanetSurveys,
    title = {{Planet populations inferred from debris discs: Insights from 178 debris systems in the ISPY, LEECH, and LIStEN planet-hunting surveys}},
    year = {2022},
    journal = {A\&A},
    author = {Pearce, Tim D. and Launhardt, Ralf and Ostermann, Robert and Kennedy, Grant M. and Gennaro, Mario and Booth, Mark and Krivov, Alexander V. and Cugno, Gabriele and Henning, Thomas K. and Quirrenbach, Andreas and Barcucci, Arianna Musso and Matthews, Elisabeth C. and Ruh, Henrik L. and Stone, Jordan M.},
    month = {3},
    volume = {659},
    publisher = {EDP Sciences},
    doi = {10.1051/0004-6361/202142720},
    issn = {14320746},
    arxivId = {2201.08369}
}

@article{Marino2020PopulationStars,
    title = {{Population synthesis of exocometary gas around A stars}},
    year = {2020},
    journal = {MNRAS},
    author = {Marino, S and Flock, M and Henning, Th and Kral, Q and Matr{\`{a}}, L and Wyatt, M C},
    number = {3},
    month = {3},
    pages = {4409--4429},
    volume = {492},
    url = {https://academic.oup.com/mnras/article/492/3/4409/5717325 http://arxiv.org/abs/2001.10543 http://dx.doi.org/10.1093/mnras/stz3487},
    doi = {10.1093/mnras/stz3487},
    issn = {0035-8711},
    arxivId = {2001.10543}
}

@article{Kral2023PotentialWinds,
    title = {{Potential effects of stellar winds on gas dynamics in debris disks leading to observable belt winds}},
    year = {2023},
    journal = {A{\&}A},
    author = {Kral, Quentin and Pringle, J. E. and Matr{\`{a}}, Luca and Th{\'{e}}bault, Philippe},
    month = {1},
    pages = {A116},
    volume = {669},
    url = {http://arxiv.org/abs/2211.04191 https://www.aanda.org/10.1051/0004-6361/202243729},
    doi = {10.1051/0004-6361/202243729},
    issn = {0004-6361},
    arxivId = {2211.04191}
}

@article{Cataldi2023PrimordialALMA,
    title = {{Primordial or Secondary? Testing Models of Debris Disk Gas with ALMA}},
    year = {2023},
    journal = {ApJ},
    author = {Cataldi, Gianni and Aikawa, Yuri and Iwasaki, Kazunari and Marino, Sebastian and Brandeker, Alexis and Hales, Antonio and Henning, Thomas and Higuchi, Aya E. and Hughes, A. Meredith and Janson, Markus and Kral, Quentin and Matr{\`{a}}, Luca and Mo{\'{o}}r, Attila and Olofsson, Göran and Redfield, Seth and Roberge, Aki},
    number = {2},
    pages = {111},
    volume = {951},
    isbn = {1342247736},
    doi = {10.3847/1538-4357/acd6f3},
    issn = {0004-637X},
    arxivId = {2305.12093}
}

@article{Bohn2022ProbingObservations,
    title = {{Probing inner and outer disk misalignments in transition disks: Constraints from VLTI/GRAVITY and ALMA observations}},
    year = {2022},
    journal = {A\&A},
    author = {Bohn, A. J. and Benisty, M. and Perraut, K. and Van Der Marel, N. and W{\"{o}}lfer, L. and Van Dishoeck, E. F. and Facchini, S. and Manara, C. F. and Teague, R. and Francis, L. and Berger, J. P. and Garcia-Lopez, R. and Ginski, C. and Henning, T. and Kenworthy, M. and Kraus, S. and M{\'{e}}nard, F. and M{\'{e}}rand, A. and P{\'{e}}rez, L. M.},
    month = {2},
    volume = {658},
    publisher = {EDP Sciences},
    doi = {10.1051/0004-6361/202142070},
    issn = {14320746}
}

@article{Hughes2017RadialCeti,
    title = {{Radial Surface Density Profiles of Gas and Dust in the Debris Disk around 49 Ceti}},
    year = {2017},
    journal = {ApJ},
    author = {Hughes, A Meredith and Lieman-Sifry, Jesse and Flaherty, Kevin M and Daley, Cail M and Roberge, Aki and K{\'{o}}sp{\'{a}}l, Ágnes and Mo{\'{o}}r, Attila and Kamp, Inga and Wilner, David J and Andrews, Sean M and Kastner, Joel H and {\'{A}}brah{\'{a}}m, Peter},
    pages = {86},
    volume = {839},
    url = {http://iopscience.iop.org/article/10.3847/1538-4357/aa6b04/pdf},
    doi = {10.3847/1538-4357/aa6b04}
}

@article{Arce-Tord2023Radio-continuum44,
    title = {{Radio-continuum decrements associated to shadowing from the central warp in transition disc DoAr 44}},
    year = {2023},
    journal = {MNRAS},
    author = {Arce-Tord, Carla and Casassus, Simon and Dent, William R.F. and Perez, Sebastian and Carcamo, Miguel and Weber, Philipp and Engler, Natalia and Cieza, Lucas A. and Hales, Antonio and Zurlo, Alice and Marino, Sebastian},
    number = {2},
    month = {12},
    pages = {2077--2085},
    volume = {526},
    publisher = {Oxford University Press},
    doi = {10.1093/mnras/stad2885},
    issn = {13652966},
    arxivId = {2309.10735}
}

@article{Lovell2021RapidDisc,
    title = {{Rapid CO gas dispersal from NO Lup's class III circumstellar disc}},
    year = {2021},
    journal = {MNRASL},
    author = {Lovell, J B and Kennedy, G M and Marino, S and Wyatt, M C and Ansdell, M and Kama, M and Manara, C F and Matr{\`{a}}, L and Rosotti, G and Tazzari, M and Testi, L and Williams, J P},
    number = {1},
    pages = {L66-L71},
    volume = {502},
    url = {http://arxiv.org/abs/2011.13229},
    doi = {10.1093/mnrasl/slaa189},
    issn = {17453933},
    arxivId = {2011.13229}
}

@article{Matra2025REsolvedWavelengths,
    title = {{REsolved ALMA and SMA Observations of Nearby Stars (REASONS): A population of 74 resolved planetesimal belts at millimetre wavelengths}},
    year = {2025},
    journal = {A\&A},
    author = {Matr{\`{a}}, L. and Marino, S. and Wilner, D. J. and Kennedy, G. M. and Booth, M. and Krivov, A. V. and Williams, J. P. and Hughes, A. M. and Del Burgo, C. and Carpenter, J. and Davies, C. L. and Ertel, S. and Kral, Q. and Lestrade, J. F. and Marshall, J. P. and Milli, J. and {\"{O}}berg, K. I. and Pawellek, N. and Sepulveda, A. G. and Wyatt, M. C. and Matthews, B. C. and Macgregor, M.},
    month = {1},
    volume = {693},
    publisher = {EDP Sciences},
    doi = {10.1051/0004-6361/202451397},
    issn = {14320746}
}

@article{Booth2016ResolvingALMA,
    title = {{Resolving the planetesimal belt of HR 8799 with ALMA}},
    year = {2016},
    journal = {MNRAS},
    author = {Booth, Mark and Jord{\'{a}}n, Andrés and Casassus, Simon and Hales, Antonio S and Dent, William R F and Faramaz, Virginie and Matr{\`{a}}, Luca and Barkats, Denis and Brahm, Rafael and Cuadra, Jorge},
    pages = {10--14},
    volume = {460},
    url = {https://help.almascience.org/index.php?/Knowledgebase},
    doi = {10.1093/mnrasl/slw040}
}

@article{Andrews2018ScalingDisks,
    title = {{Scaling Relations Associated with Millimeter Continuum Sizes in Protoplanetary Disks}},
    year = {2018},
    journal = {ApJ},
    author = {Andrews, Sean M. and Terrell, Marie and Tripathi, Anjali and Ansdell, Megan and Williams, Jonathan P. and Wilner, David J.},
    number = {2},
    month = {10},
    pages = {157},
    volume = {865},
    url = {https://iopscience.iop.org/article/10.3847/1538-4357/aadd9f},
    doi = {10.3847/1538-4357/aadd9f},
    issn = {0004-637X}
}

@article{Krivov2021SolutionSmall,
    title = {{Solution to the debris disc mass problem: Planetesimals are born small?}},
    year = {2021},
    journal = {MNRAS},
    author = {Krivov, Alexander V. and Wyatt, Mark C.},
    number = {1},
    pages = {718--735},
    volume = {500},
    publisher = {Oxford University Press},
    doi = {10.1093/mnras/staa2385},
    issn = {13652966},
    arxivId = {2008.07406}
}

@article{Gaspar2023SpatiallyJWST/MIRI,
    title = {{Spatially resolved imaging of the inner Fomalhaut disk using JWST/MIRI}},
    year = {2023},
    journal = {Nature Ast},
    author = {G{\'{a}}sp{\'{a}}r, András and Wolff, Schuyler Grace and Rieke, George H. and Leisenring, Jarron M. and Morrison, Jane and Su, Kate Y. L. and Ward-Duong, Kimberly and Aguilar, Jonathan and Ygouf, Marie and Beichman, Charles and Llop-Sayson, Jorge and Bryden, Geoffrey},
    month = {5},
    url = {http://arxiv.org/abs/2305.03789%0Ahttp://dx.doi.org/10.1038/s41550-023-01962-6 https://www.nature.com/articles/s41550-023-01962-6},
    doi = {10.1038/s41550-023-01962-6},
    issn = {2397-3366},
    arxivId = {2305.03789}
}

@article{Pericaud2017,
    title = {{The hybrid disks: a search and study to better understand evolution of disks}},
    year = {2017},
    journal = {A{\&}A},
    author = {P{\'{e}}ricaud, J and Di Folco, E and Dutrey, A and Guilloteau, S and Pi{\'{e}}tu, V},
    month = {4},
    pages = {A62},
    volume = {600},
    url = {https://www.aanda.org/articles/aa/pdf/2017/04/aa29371-16.pdf http://www.aanda.org/10.1051/0004-6361/201629371},
    doi = {10.1051/0004-6361/201629371},
    issn = {0004-6361}
}

@article{Holland1998SubmillimetreStars,
    title = {{Submillimetre images of dusty debris around nearby stars}},
    year = {1998},
    journal = {Nature},
    author = {Holland, Wayne S and Greaves, Jane S and Zuckerman, B and Webb, R A and Mccarthy, Chris and Coulson, Iain M and Walther, D M and Dent, William R F and Geark, Walter K and Robson, Ian},
    pages = {788},
    volume = {392}
}

@article{TheAstropyCollaboration2018ThePackage,
    title = {{The Astropy Project: Building an inclusive, open-science project and status of the v2.0 core package}},
    year = {2018},
    author = {{The Astropy Collaboration} and Price-Whelan, A M and Sip{\H{o}}cz, B M and G{\"{u}}nther, H M and Lim, P L and Crawford, S M and Conseil, S and Shupe, D L and Craig, M W and Dencheva, N and Ginsburg, A and VanderPlas, J T and Bradley, L D and P{\'{e}}rez-Su{\'{a}}rez, D and de Val-Borro, M and Aldcroft, T L and Cruz, K L and Robitaille, T P and Tollerud, E J and Ardelean, C and Babej, T and Bachetti, M and Bakanov, A V and Bamford, S P and Barentsen, G and Barmby, P and Baumbach, A and Berry, K L and Biscani, F and Boquien, M and Bostroem, K A and Bouma, L G and Brammer, G B and Bray, E M and Breytenbach, H and Buddelmeijer, H and Burke, D J and Calderone, G and Rodr{\'{i}}guez, J L Cano and Cara, M and Cardoso, J V M and Cheedella, S and Copin, Y and Crichton, D and D{\'{A}}vella, D and Deil, C and Depagne, É and Dietrich, J P and Donath, A and Droettboom, M and Earl, N and Erben, T and Fabbro, S and Ferreira, L A and Finethy, T and Fox, R T and Garrison, L H and Gibbons, S L J and Goldstein, D A and Gommers, R and Greco, J P and Greenfield, P and Groener, A M and Grollier, F and Hagen, A and Hirst, P and Homeier, D and Horton, A J and Hosseinzadeh, G and Hu, L and Hunkeler, J S and Ivezi{\'{c}}, Ž and Jain, A and Jenness, T and Kanarek, G and Kendrew, S and Kern, N S and Kerzendorf, W E and Khvalko, A and King, J and Kirkby, D and Kulkarni, A M and Kumar, A and Lee, A and Lenz, D and Littlefair, S P and Ma, Z and Macleod, D M and Mastropietro, M and McCully, C and Montagnac, S and Morris, B M and Mueller, M and Mumford, S J and Muna, D and Murphy, N A and Nelson, S and Nguyen, G H and Ninan, J P and N{\"{o}}the, M and Ogaz, S and Oh, S and Parejko, J K and Parley, N and Pascual, S and Patil, R and Patil, A A and Plunkett, A L and Prochaska, J X and Rastogi, T and Janga, V Reddy and Sabater, J and Sakurikar, P and Seifert, M and Sherbert, L E and Sherwood-Taylor, H and Shih, A Y and Sick, J and Silbiger, M T and Singanamalla, S and Singer, L P and Sladen, P H and Sooley, K A and Sornarajah, S and Streicher, O and Teuben, P and Thomas, S W and Tremblay, G R and Turner, J E H and Terr{\'{o}}n, V and van Kerkwijk, M H and de la Vega, A and Watkins, L L and Weaver, B A and Whitmore, J B and Woillez, J and Zabalza, V},
    month = {10},
    url = {http://arxiv.org/abs/1801.02634 http://dx.doi.org/10.3847/1538-3881/aabc4f},
    doi = {10.3847/1538-3881/aabc4f},
    journal = {arXiv e-prints},
    volume = {1801.02634},
    arxivId = {1801.02634}
}

@article{Eilers2019TheKpc,
    title = {{The Circular Velocity Curve of the Milky Way from 5 to 25 kpc}},
    year = {2019},
    journal = {ApJ},
    author = {Eilers, Anna-Christina and Hogg, David W. and Rix, Hans-Walter and Ness, Melissa K.},
    number = {1},
    month = {1},
    pages = {120},
    volume = {871},
    publisher = {IOP Publishing},
    url = {https://iopscience.iop.org/article/10.3847/1538-4357/aaf648 https://iopscience.iop.org/article/10.3847/1538-4357/aaf648/meta},
    doi = {10.3847/1538-4357/AAF648},
    issn = {0004-637X},
    arxivId = {1810.09466}
}
\bibliographystyle{aa.bst}

\begin{appendix}

\section{Model fit to the continuum visibilities}
\label{sec:vis_modelfit}

We employ Monte Carlo Markov Chain sampling, through the package emcee \citep{Foreman-Mackey2013EmceeHammer}, to sample the parameter space whilst applying a Gaussian ring model, as described by Equation \ref{eqn:gaussRing}, to the de-projected observed visibilities. The representative model, which takes the median values (50th percentile), is used to construct the model included in Figure \ref{fig:tclean_images} and the subsequent residual map.

\begin{figure}
    \centering
    \includegraphics[width=0.8\linewidth]{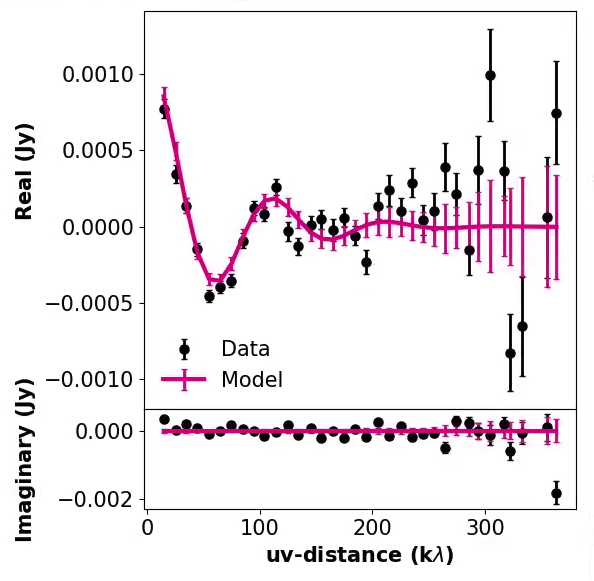} 
    \label{fig:vis_fit}
\end{figure}

\begin{figure}
    \centering
    \includegraphics[width=0.8\linewidth]{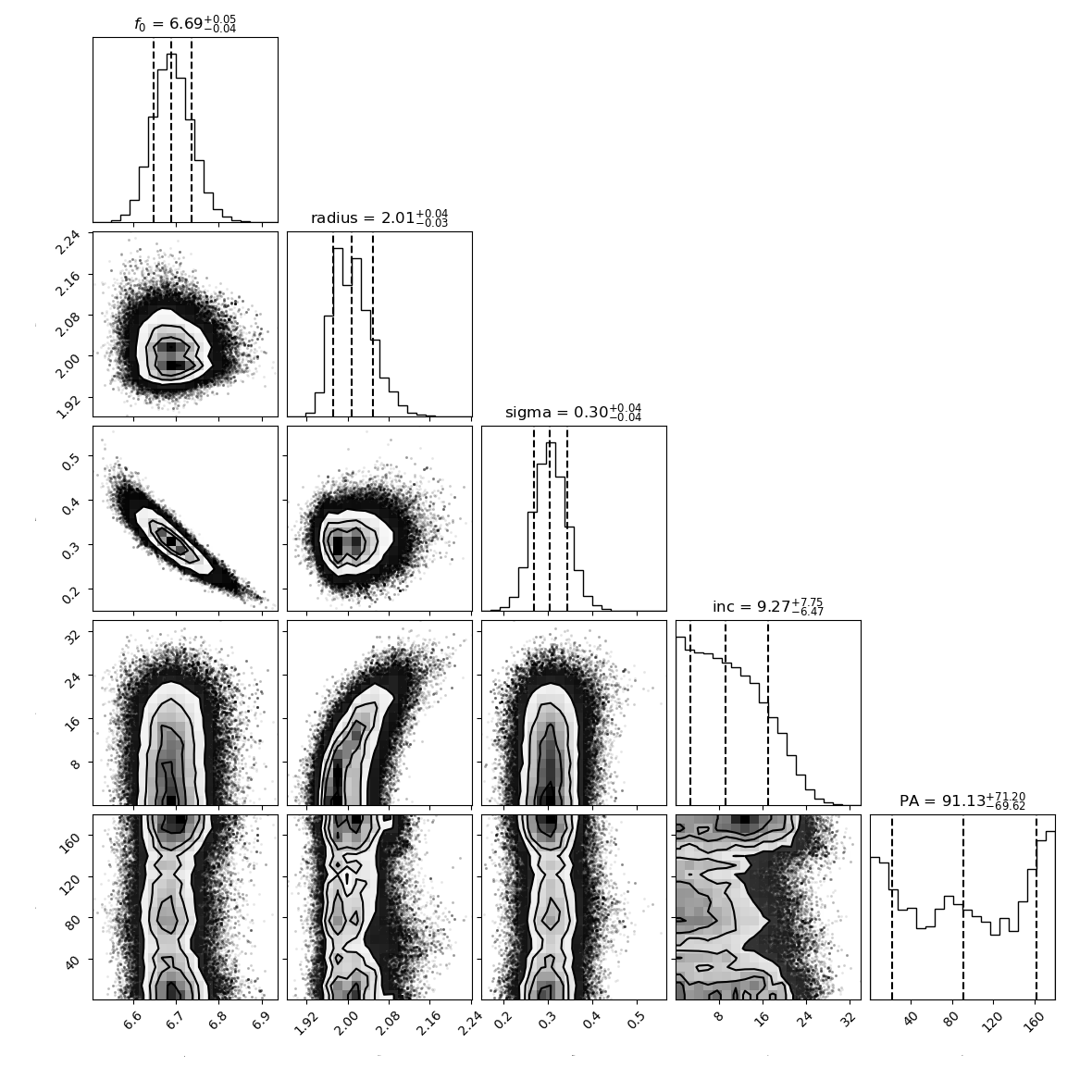} 
    \caption{\textbf{Top: } Comparison of the real and imaginary components of the representative model as a function of uv distance with the observed visibilities in Fourier space. Error bars show the standard deviation of the the visibilities in each radial bin. \textbf{Bottom: }Corner plot of the posterior distributions from the fit of our brightness distribution model to the observed visibilities. The values adopted for our representative model are given in the titles for each of the parameters as the 50th percentile of the posterior distribution, with error bars taken as the 16th and 84th percentile.  }
    \label{fig:vis_fit}
\end{figure}

\section{Channel maps from the $^{12}$CO cube}
\label{sec:appendix12CO}

\begin{figure}
    \centering
    \includegraphics[width=0.8\linewidth]{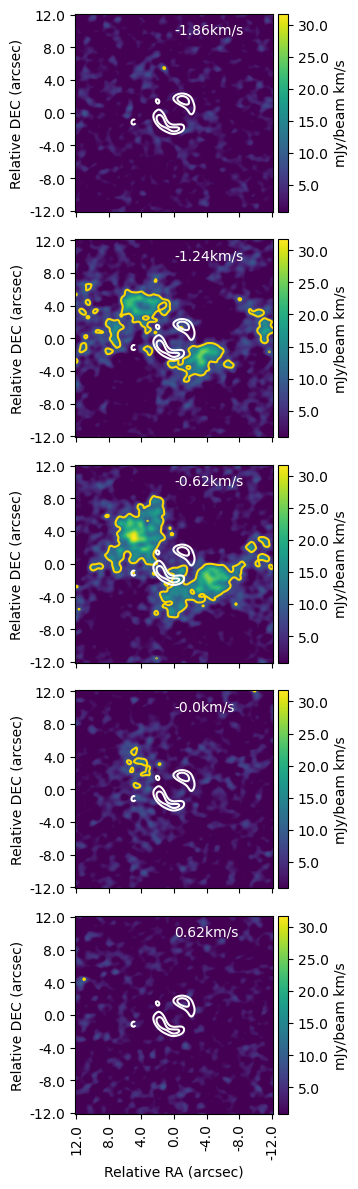}
    \caption{Channels maps showing the $^{12}$CO emission. The white lines show the continuum emission from the image where a uvtaper was applied, plotted at contours of 3,4$ \times \sigma_{\rm cont}$. Yellow contours show the 5$ \times \sigma_{\rm CO}$ level in individual channels.}
    \label{fig:line_cube}
\end{figure}

\end{appendix}

\end{document}